\documentclass[10pt,journal,twocolumn,twoside]{IEEEtran} 
\pdfoutput=1 
\usepackage{cite}
\usepackage{amsmath,amssymb,amsfonts}
\usepackage{algorithm,algpseudocode}
\usepackage{color}
\usepackage{multirow}
\usepackage{booktabs}
\usepackage{mathtools}

\usepackage{fancyhdr}
\usepackage{xcolor}

\usepackage{hyperref}
\hypersetup{
	pdftitle={Schiller2021 - Suboptimal nonlinear moving horizon estimation},
	pdfauthor={Julian D. Schiller, Matthias A. Mueller},
	colorlinks,
	linkcolor={red!50!black},
	citecolor={blue!50!black},
	urlcolor={blue!80!black}
}

\def\doi{10.1109/TAC.2022.3173937}
\fancyfootoffset{-1em}
\fancypagestyle{copyright}{
	\cfoot{\vspace{-1.8em}\footnotesize\textcopyright{} 2022 IEEE. Personal use of this material is permitted.  Permission from IEEE must be obtained for all other uses, in any current or future media, including reprinting/republishing this material for advertising or promotional purposes, creating new collective works, for resale or redistribution to servers or lists, or reuse of any copyrighted component of this work in other works.}
	\chead{\footnotesize This article has been accepted for publication in IEEE Transactions on Automatic Control. This is the author's version which has not been fully edited and content may change prior to final publication. Citation information: \href{http://dx.doi.org/\doi}{DOI \doi}}
}

\usepackage{amsthm}
\newtheorem{thm}{Theorem}
\newtheorem{lem}{Lemma}

\newtheorem{cor}{Corollary}
\newtheorem{defn}{Definition}
\newtheorem{rem}{Remark}
\newtheorem{assum}{Assumption}

\begin{document}
	
\title{Suboptimal nonlinear moving horizon estimation}
\author{Julian D. Schiller, \IEEEmembership{Graduate Student Member, IEEE}, and Matthias A. Müller, \IEEEmembership{Senior Member, IEEE}
\thanks{This work was supported by the German Research Foundation~(DFG) under the research grant MU-3929-2/1. The material in this paper was partially presented at the 7th IFAC Conference on Nonlinear Model Predictive Control (NMPC), July 11-14, 2021, Bratislava, Slovakia. \textit{(Corresponding author: Julian D. Schiller.)} %
}
\thanks{The authors are with the Institute of Automatic Control, Leibniz University Hannover, Germany (e-mail: {\scriptsize$\{$}schiller,mueller{\scriptsize$\}$}@irt.uni-hannover.de). }
}

\maketitle
\thispagestyle{copyright}

\begin{abstract}
	In this paper, we propose a suboptimal moving horizon estimator for a general class of nonlinear systems.
	For the stability analysis, we transfer the ``feasibility-implies-stability/robustness'' paradigm from model predictive control to the context of moving horizon estimation in the following sense: 
	Using a suitably defined, feasible candidate solution based on an auxiliary observer, robust stability of the proposed suboptimal estimator is inherited independently of the horizon length and even if no optimization is performed.
	Moreover, the proposed design allows for the choice between two cost functions different in structure:
	the former in the manner of a standard least squares approach, which is typically used in practice, and the latter following a time-discounted modification, resulting in better theoretical guarantees.
	We apply the proposed suboptimal estimator to {a nonlinear chemical reactor process, verify} the theoretical assumptions, {and} show that even a few iterations of the optimizer are sufficient to significantly improve the estimation results of the auxiliary observer. 
	{Furthermore, we illustrate the flexibility of the proposed design by employing different solvers and compare the performance with two state-of-the-art fast MHE schemes from the literature.}
\end{abstract}


\section{Introduction}\label{sec:introduction}
	Knowledge of the internal state of a dynamical system is crucial for many control applications, e.g., for stabilizing the system via state feedback or for monitoring compliance with safety-critical conditions.
	In most practical cases, however, the state cannot be completely measured and therefore must be reconstructed using the (measurable) system output.
	This is particularly challenging if nonlinear systems with constraints are present and robustness against model inaccuracies and measurement noise is to be ensured.
	To this end, moving horizon estimation (MHE) has proven to be a powerful solution to the state estimation problem and various theoretical guarantees such as robust stability properties have been established in recent years, see, e.g., \cite{Rawlings2017,Knuefer2021,Hu2021,Mueller2017,Allan2019a,Allan2020a}.
	In MHE, the current state is estimated by optimizing over a fixed number of past measurements, taking into account both system dynamics and constrained sets of decision variables. 
	However, since this approach requires solving a usually non-convex optimization problem at each time step, MHE is computationally demanding.
	Moreover, since the computing power available in practice is often severely limited, optimization-based techniques usually can only be applied to systems with a fairly large sampling interval.

	\subsubsection*{Related work}
	In order to make the applicability of MHE in real-time more likely, methods based on an additional \textit{auxiliary observer} were developed, among others, thus simplifying the optimization problem significantly.
	For example, in \cite{Sui2010}, a pre-estimating MHE scheme for linear systems was proposed that utilized an additional observer to replace the state equation as a dynamical constraint.
	Since this allows to compensate for model uncertainties without computing an optimal disturbance sequence, the optimization variables could be reduced to one, namely the initial state at the beginning of the horizon.
	In~\cite{Suwantong2014}, this idea was transferred to a class of nonlinear systems, and a major speed improvement compared to standard MHE could be shown.
	However, this results in a loss of degrees of freedom, since there is no possibility to weight model disturbances and measurement noise differently in the optimization problem.
	In \cite{Liu2013}, an observer was employed to construct a confidence region for the actual system state.
	Introducing this region as an additional constraint in the optimization problem can, however, be quite restrictive and hence might not allow for major improvements of MHE compared to the auxiliary observer.
	In \cite{Gharbi2020b}, a proximity-MHE scheme was proposed for a general class of nonlinear systems, where an additional observer is used to construct a stabilizing a priori estimate yielding a proper warm start for the optimization problem, and nominal stability could be shown by Lyapunov arguments.
	
	However, all the above methods still require optimal solutions to the MHE problem, and their complete computation within fixed time intervals is difficult (if not impossible) to guarantee.
	A more intuitive approach is to simply terminate the underlying optimization algorithm after a fixed number of iterations, which on the one hand provides only suboptimal estimates, but on the other hand ensures fixed computation times.
	However, since most results from the nonlinear MHE literature are crucially based on optimality~\cite{Rawlings2017,Mueller2017,Hu2021,Knuefer2021,Allan2019a,Allan2020a}, stability of suboptimal MHE cannot be straightforwardly deduced.
	For practical (real-time) applications, it is therefore crucial to develop suboptimal schemes that guarantee robust stability without requiring optimal solutions.
	Nevertheless, there are some fast (real-time) MHE schemes available in the literature that are based on specific optimization algorithms, e.g., utilizing gradient, conjugate-gradient or Newton methods~\cite{Kuehl2011,Wynn2014,Alessandri2017,Wan2017}.
	However, the corresponding results rely on (local) contraction properties of the specific algorithms and therefore require both a proper initial guess and at least one iteration to ensure (local) stability.
	{In \cite{Gharbi2020a}, a suboptimal proximity-MHE scheme for linear systems was proposed, where nominal stability guarantees could be given without performing any optimization by using a pre-stabilizing observer and contraction properties of a certain gradient-based optimization algorithm.
	This approach has recently been extended to nonlinear systems in \cite{Gharbi2021}, thus providing nominal stability guarantees for a suboptimal nonlinear proximity-MHE scheme, relying on local properties of the optimization algorithm involved.}

	\subsubsection*{Contribution}
	In this paper, we establish the ``fea\-si\-bil\-i\-ty-im\-plies-sta\-bil\-i\-ty/ro\-bust\-ness'' paradigm from model predictive control (MPC) in the context of nonlinear MHE.
	Indeed, it is well known that if the suboptimal solution to a given MPC problem can be guaranteed to improve the cost of a suitably chosen warm start, then robust stability of the controller can be directly inferred \cite{Scokaert1999,Pannocchia2011a}.
	Transferring this concept to nonlinear MHE, we prove robust stability of the proposed suboptimal estimator (i) regardless of the chosen length of the estimation horizon and (ii) without explicitly performing any optimization by simply requiring that a suboptimal solution to a given MHE problem improves the cost of a feasible \textit{candidate solution}.
	To this end, we propose two different candidate solutions applicable to different nonlinear system classes, both of which rely on an additional, robustly exponentially stable auxiliary observer (cf. Sections~\ref{sec:candidate_nominal} and~\ref{sec:candidate_observer}).    
	Moreover, we consider two different cost functions, thus allowing for the choice between a standard, commonly chosen least squares type \cite{Allan2019a,Ji2016,Mueller2017,Rawlings2017,Allan2020a}, and a time-discounted approach motivated by \cite{Knuefer2018,Knuefer2021}, which leads to better theoretical guarantees.
	By employing a re-initialization strategy, we also enable the auxiliary observer to incorporate improved suboptimal estimates from the past.
	Assuming nonlinear exponentially detectable systems \cite{Knuefer2018,Knuefer2020,Allan2021}, we provide explicit formulas for the (robust) bound on the suboptimal estimation error for different configurations of the proposed estimator and compare their main characteristics and requirements (cf. Section~\ref{sec:discussion}).
	Furthermore, we provide the ability to include state and output constraints into the suboptimal MHE scheme, even if they are violated by the auxiliary observer (cf. Section~\ref{sec:constraints}).
	A preliminary version of this framework for the special case of additive disturbances using the standard (non-discounted) cost function, without re-initialization of the auxiliary observer and without explicit constraint handling was presented at a conference~\cite{Schiller2021}. 
	
	Overall, the proposed suboptimal MHE framework is applicable to a large class of nonlinear systems and guarantees robust stability in case of unknown disturbances and noise. 
	{We illustrate the theoretical results in terms of an extensive simulation case study involving a nonlinear chemical reactor process (cf. Section~\ref{sec:simulation}).}
	After numerically verifying the theoretical assumptions (in particular, robust exponential stability of the auxiliary observer and exponential detectability of the nonlinear system), we show that performing only a few iterations of the optimizer each time step is already sufficient to significantly improve the estimation results of the auxiliary observer.
	{In addition, we provide a numerical comparison of our methods with the fast MHE schemes from \cite{Kuehl2011,Wynn2014}, where we achieve comparable performance both in terms of accuracy and required computation time.}
	
	\subsubsection*{Notation}
	Let the set of all integers in an interval $[a,b] \subset \mathbb{R}$ be denoted by $\mathbb{I}_{[a,b]}$ and the set of all integers greater than or equal to $a$ by $\mathbb{I}_{\geq a}$.
	We define $|x|$ to be the Euclidean norm of the vector $x \in \mathbb{R}^n$. 
	Symbols in bold type represent sequences of vectors, i.e. $\boldsymbol{x} = \lbrace x(0), x(1),...\rbrace$, which can be either of finite length (e.g., of length $K$ for some $K \in \mathbb{I}_{\geq1}$, denoted by $\boldsymbol{x} \in\mathbb{X}^{K}$), or of infinite length (denoted by $\boldsymbol{x} \in\mathbb{X}^{\infty}$).

\section{Preliminaries and setup}\label{sec:setup}

\subsection{System description}
We consider the discrete-time, perturbed nonlinear system
\begin{subequations} \label{eq:system}
	\begin{flalign} \label{eq:system_a}
	x_{t+1} &= f(x_t,u_t,w_t), \\ 
	y_t &= h(x_t,u_t,v_t), \label{eq:system_b}                                                                                                                                                                                               
	\end{flalign}
\end{subequations}                                                  
with time $t \in \mathbb{I}_{\geq0}$, and where $x \in \mathbb{X} = \mathbb{R}^{n_x}$ is the system state, $u \in \mathbb{U} \subseteq  \mathbb{R}^{n_u}$ is the (known) control input, $w \in \mathbb{W} \subseteq \mathbb{R}^{n_w}$ is the (unknown) process disturbance, $v \in \mathbb{V} \subseteq  \mathbb{R}^{n_v}$ is the (unknown) measurement noise, and $y \in \mathbb{Y} = \mathbb{R}^{n_y}$ is the measured output.
We treat $\mathbb{U}$, $\mathbb{W}$, and $\mathbb{V}$ as known sets that are inherently fulfilled by the original system and assume that $0 \in \mathbb{W}$ and $0 \in \mathbb{V}$.
Note that the former assumptions on the domain of the system are modified in Sections~\ref{sec:candidate_observer}-\ref{sec:constraints}, where we consider the special case of additive disturbances and state and output constraints.                                                                                                                                                                                                                                  
The mappings $f:\mathbb{X} \times \mathbb{U} \times \mathbb{W} \rightarrow \mathbb{X}$ and $h: \mathbb{X} \times \mathbb{U} \times \mathbb{V} \rightarrow \mathbb{Y}$ are some nonlinear continuous functions representing the system dynamics and output model, respectively, and their corresponding nominal equations are denoted as $f_n(x,u) = f(x,u,0)$ and $h_n(x,u) = h(x,u,0)$.
We additionally impose the following continuity property on~$h$.

\begin{assum} \label{ass:lipschitz} 
	The function $h$ is Lipschitz continuous, i.e., there exists some constant $H > 0$ such that
	$|h(x,u,v)-h(\chi,\mu,\nu)|\leq H(|x-\chi|+|u-\mu|+|v-\nu|)$ for all $x,\chi \in \mathbb{X}$, $u,\mu \in \mathbb{U}$, and $v,\nu \in \mathbb{V}$.
\end{assum} 

An initial state $x_0 \in \mathbb{X}$ together with the input sequences $\boldsymbol{u}=\lbrace u_0,u_{1},\ldots \rbrace  \in \mathbb{U}^{\infty}$ and the disturbance sequences $\boldsymbol{w}=\lbrace w_0,w_{1},\ldots \rbrace  \in \mathbb{W}^{\infty}$ and $\boldsymbol{v}=\lbrace v_0,v_{1},\ldots \rbrace  \in \mathbb{V}^{\infty}$ yields a state sequence $\boldsymbol{x}=\lbrace x_0,x_{1}, \ldots \rbrace  \in \mathbb{X}^{\infty}$ and an output sequence $\boldsymbol{y}=\lbrace y_0,y_{1},\ldots \rbrace  \in \mathbb{Y}^{\infty}$ under (\ref{eq:system}).
In the following, we call $(\boldsymbol{x}, \boldsymbol{u}, \boldsymbol{w}, \boldsymbol{v}, \boldsymbol{y}) \in \Sigma$ a trajectory of system (\ref{eq:system}) and $\Sigma \subset \mathbb{X}^{\infty} \times \mathbb{U}^{\infty} \times \mathbb{W}^{\infty} \times \mathbb{V}^{\infty} \times \mathbb{Y}^{\infty}$ the set of all possible trajectories that satisfy (\ref{eq:system}) for all $t \in \mathbb{I}_{\geq0}$.

{In order to establish robust stability of the proposed suboptimal estimator, an appropriate detectability assumption is required. To this end, we consider the following notion of exponential incremental input/output-to-state stability (e-IOSS).}

\begin{defn}[e-IOSS] \label{def:eIOSS} System (\ref{eq:system}) is  e-IOSS if there exist constants $c_p,c_u,c_w,c_v,c_y>0$ and $\eta \in (0,1)$ such that any two trajectories of (\ref{eq:system}) given by $(\boldsymbol{x}, \boldsymbol{u}, \boldsymbol{w}, \boldsymbol{v}, \boldsymbol{y}), (\boldsymbol{\chi}, \boldsymbol{\mu}, \boldsymbol{\omega}, \boldsymbol{\nu}, \boldsymbol{\zeta}) \in \Sigma$ satisfy, for all $t \in \mathbb{I}_{\geq0}$,
	\begin{align} 
	&| x_t - \chi_t | \leq  c_p|x_0 - \chi_0|\eta^t \label{eq:eIOSS} + \sum_{\tau=1}^t\eta^{\tau} 
	\big(c_u|u_{t-\tau} - \mu_{t-\tau}| \\
	& \
	+ c_w|w_{t-\tau} - \omega_{t-\tau}|  + c_v|v_{t-\tau} - \nu_{t-\tau}| 
	+ c_y|y_{t-\tau} - \zeta_{t-\tau}|\big). \notag
	\end{align}
\end{defn}

{Definition \ref{def:eIOSS} is an exponential version of incremental input/output-to-state stability (i-IOSS), which became standard as a notion of nonlinear detectability in the context of MHE in recent years\footnote{
	{Note that the asymptotic-gain formulation of i-IOSS (cf. \cite{Mueller2017,Allan2019a,Ji2016}) is equivalent (cf. \cite[Prop. 2.5]{Allan2021}) to the time-discounted formulation used in, e.g., \cite{Rawlings2017,Knuefer2020,Allan2021,Allan2020a,Hu2021,Knuefer2021,Knuefer2018}.}}
\cite{Rawlings2017,Mueller2017,Allan2019a,Ji2016,Knuefer2018,Knuefer2020,Allan2021,Allan2020a,Hu2021,Knuefer2021}.
To account for the general nonlinear setup~\eqref{eq:system}, we  treat the influences of the inputs, outputs, and their respective (nonlinear) disturbances individually in~\eqref{eq:eIOSS} as suggested in \cite{Knuefer2020,Knuefer2021}, where we additionally consider an exponential decay in order to simplify the subsequent analysis.
We point out that the recent results on i-IOSS from \cite{Knuefer2020,Allan2021} can be straightforwardly formulated for the special case of exponential stability, and one can show that e-IOSS is both necessary (compare \cite[Prop. 3]{Knuefer2020}, \cite[Prop. 2.6]{Allan2021}) and sufficient (compare \cite[Thm. 14, Rem. 17]{Knuefer2021}) for the existence of robustly exponentially stable state estimators.
Moreover, e-IOSS can be verified using a suitable Lyapunov function (compare \cite{Allan2021,Knuefer2020}), which can be constructed based on the differential dynamics as suggested in \cite[Sec.~4]{Schiller2022} or using an additional robustly exponentially stable observer, compare also \cite[Prop.~4]{Koehler2021} and the simulation example in Section~\ref{sec:simulation}.}

\subsection{Suboptimal moving horizon estimator}\label{sec:MHE}
The moving horizon estimator for system (\ref{eq:system}) considers at each time $t \in \mathbb{I}_{\geq0}$ past input and output data in a moving time window of length $\mathcal{N} = \min \lbrace t, N \rbrace$ for some fixed $N \in \mathbb{I}_{\geq 1}$.
Given the corresponding input and output sequences $\boldsymbol{u}_t = \lbrace u_{t-\mathcal{N}}, \ldots, u_{t-1} \rbrace$ and $\boldsymbol{y}_t = \lbrace y_{t-\mathcal{N}}, \ldots, y_{t-1} \rbrace$, the moving horizon estimate for system (\ref{eq:system}) at time $t \in \mathbb{I}_{\geq 0}$ corresponds to the minimizer of
\begin{equation} \label{eq:optiMHE}
\min_{\chi_{t-\mathcal{N}|t}, \boldsymbol{\omega}_t, \boldsymbol{\nu}_t}  J(\chi_{t-\mathcal{N}|t}, \boldsymbol{\omega}_t, \boldsymbol{\nu}_t)
\end{equation}
subject to
\begin{equation} \label{eq:constraints}
(\boldsymbol{\chi}_t, \boldsymbol{u}_t, \boldsymbol{\omega}_t, \boldsymbol{\nu}_t, \boldsymbol{\zeta}_t ) \in \Sigma^{\mathcal{N}},
\end{equation}
where  $J: \mathbb{X} \times \mathbb{W}^{\mathcal{N}} \times \mathbb{V}^{\mathcal{N}} \rightarrow \mathbb{R}_{\geq0}$ {is the cost function specified below}.
The sequences $\boldsymbol{\chi}_{t}= \lbrace \chi_{t-\mathcal{N}|t}, \ldots, \chi_{t-1|t} \rbrace \in \mathbb{X}^{\mathcal{N}}$, $\boldsymbol{\omega}_{t}= \lbrace \omega_{t-\mathcal{N}|t}, \ldots, \omega_{t-1|t} \rbrace \in \mathbb{W}^{\mathcal{N}}$, $\boldsymbol{\nu}_t= \lbrace \nu_{t-\mathcal{N}|t}, \ldots, \nu_{t-1|t} \rbrace \in \mathbb{V}^{\mathcal{N}}$, and $\boldsymbol{\zeta}_t= \lbrace \zeta_{t-\mathcal{N}|t}, \ldots, \zeta_{t-1|t} \rbrace \in \mathbb{Y}^{\mathcal{N}}$ 
contain estimates of the state, the process and measurement noise, and the corresponding system output for the time interval $\mathbb{I}_{[t-\mathcal{N},t-1]}$, estimated at time $t$,
and $\Sigma^{\mathcal{N}}$ denotes the set of all trajectories of (\ref{eq:system}) of length $\mathcal{N}$ in the time interval $\mathbb{I}_{[t-\mathcal{N},t-1]}$. 
Note that each trajectory is uniquely defined by the input sequence $\boldsymbol{u}_t$ and the decision variables $(\chi_{t-\mathcal{N}|t}, \boldsymbol{\omega}_t, \boldsymbol{\nu}_t)$ under (\ref{eq:system}).

\begin{rem}\label{rem:output_constraints}
	In the MHE literature, the estimated output $\zeta_{t-i|t} = h(\chi_{t-i|t},u_{t-i},\nu_{t-i|t})$ is usually restricted to exactly match the measured output of the real system $y_{t-i}$ for each $i \in \mathbb{I}_{[1,\mathcal{N}]}$ by imposing $\boldsymbol{\zeta}_t = \boldsymbol{y}_t$ as an additional constraint in~(\ref{eq:constraints}), see, e.g., \cite{Rawlings2017,Knuefer2021,Hu2021,Mueller2017,Allan2019a,Allan2020a}.
	However, to ensure feasibility of the candidate solutions that will be introduced in Sections~\ref{sec:candidate_nominal} and~\ref{sec:candidate_observer}, we relax this constraint and explicitly allow for different outputs.
	As a result, an additional term appears in the cost function that takes this deviation into account, compare also \cite[Rem. 8]{Knuefer2021}.
	{Note that such a formulation is generally beneficial in practice when model inaccuracies might exist, as potential infeasibility issues arising from enforcing this hard output constraint can be naturally avoided.}
\end{rem}

In analogy to \cite{Allan2019a,Mueller2017,Rawlings2017,Allan2020a}, we choose the cost function $J$ in~(\ref{eq:optiMHE}) as follows.

\begin{defn}[Non-discounted cost function] \label{def:costfunction}
	Let $t \in \mathbb{I}_{\geq 0}$, $N\in \mathbb{I}_{\geq1}$, some prior $\bar{x}_{t-\mathcal{N}}  \in \mathbb{X}$ and the input and output sequences $\boldsymbol{u}_t$ and $\boldsymbol{y}_t$ of system (\ref{eq:system}) in the time interval $\mathbb{I}_{[t-\mathcal{N},t-1]}$ be given and let $\Gamma: \mathbb{X} \times \mathbb{X} \rightarrow \mathbb{R}_{\geq0}$ and $l: \mathbb{W} \times \mathbb{V} \times \mathbb{R}^{n_y} \rightarrow \mathbb{R}_{\geq0}$.
	For $\chi_{t-\mathcal{N}|t} \in \mathbb{X}$, $\boldsymbol{\omega}_t \in \mathbb{W}^\mathcal{N}$ and $\boldsymbol{\nu}_t \in \mathbb{V}^\mathcal{N}$,
	define
	\begin{align} \label{eq:costfunction}
	J_{\text{nd}}(\chi_{t-\mathcal{N}|t}, \boldsymbol{\omega}_t, \boldsymbol{\nu}_t) &:= \Gamma(\chi_{t-\mathcal{N}|t},\bar{x}_{t-\mathcal{N}}) \\
	& \ \ \ \ + \sum_{i=1}^{\mathcal{N}}
	l(\omega_{t-i|t},\nu_{t-i|t},y_{t-i}-\zeta_{t-i|t}). \notag
	\end{align}
\end{defn}

We impose the following assumption on the cost function.

\begin{assum} \label{ass:cost}
	The functions $\Gamma$ and $l$ from Definition~\ref{def:costfunction} are continuous and satisfy for any given $y \in \mathbb{Y}$
	\begin{subequations} \label{eq:cost}
		\begin{gather} \label{eq:cost_G}
		\underline{c}_p|\chi-\bar{x}|^{a} \leq \Gamma(\chi,\bar{x}) 
		\leq \overline{c}_p|\chi-\bar{x}|^a, \ \ \ \ \ \ \\
		\underline{c}_w|\omega|^a + \underline{c}_v|\nu|^a + \underline{c}_y|y-\zeta|^a
		\leq l(\omega,\nu,y-\zeta) \label{eq:cost_l} \\
		\leq  \overline{c}_w|\omega|^a + \overline{c}_v|\nu|^a + \overline{c}_y|y-\zeta|^a \notag
		\end{gather}
	\end{subequations}
	for all $\chi, \bar{x} \in \mathbb{X}$, $\omega \in \mathbb{W}$, $\nu \in \mathbb{V}$, and some constants $\underline{c}_p, \overline{c}_p, \underline{c}_w, \overline{c}_w, \underline{c}_v, \overline{c}_v, \underline{c}_y, \overline{c}_y>0$ and $a\geq1$, and such that
	\begin{equation}
	c_i \leq \underline{c}_i, \ i \in \lbrace p, w, v, y \rbrace \label{eq:cost_lb}
	\end{equation}
	with $c_i$ from (\ref{eq:eIOSS}).
\end{assum}
                                                                                                                                                                                                                     
\begin{rem}\label{rem:cost}
	Condition (\ref{eq:cost}) requires an exponentially bounded, positive definite cost function.                          
	The additional constraint on the exponent~$a$ ensures that this function is also convex, which allows for less restrictive bounds on the estimation error compared to existing results from the literature, such as \cite{Mueller2017,Allan2019a}. 
	Note that this assumption is not overly restrictive, since it still allows for the practical relevant case of quadratic cost functions, where $a=2$. 
	{Similar to \cite[Ass.~1]{Knuefer2021}, inequality} (\ref{eq:cost_lb}) states a compatibility condition between the cost function~\eqref{eq:costfunction} and the e-IOSS definition~(\ref{eq:eIOSS}) (which can be relaxed at the expense of a slightly worse bound on the estimation error by introducing an additional factor
	$\max\lbrace1,\max_{i\in\lbrace p,w,v,y\rbrace}{c_i}/\min_{i\in\lbrace p,w,v,y\rbrace}{\underline{c}_i}\rbrace \geq 1$).
	{We note that, in general, better guarantees for suboptimal MHE result for smaller values of $\overline{c}_y$ (and $\overline{c}_w$), cf.~Table~\ref{tab:MHE} below.
	Since these are bounded from below by~\eqref{eq:cost_lb} via~\eqref{eq:cost_l}, parameterizing the cost function~\eqref{eq:costfunction} such that the compatibility condition~\eqref{eq:cost_lb} is satisfied with equality leads to the best possible guarantees, compare also \cite[Rem.~6]{Knuefer2021} and the simulation example in Section~\ref{sec:simulation}; a similar statement applies to Assumption~\ref{ass:cost_td} below.}
\end{rem}

Definition~\ref{def:costfunction} corresponds to the type of cost function that is traditionally chosen in the nonlinear MHE literature \cite{Allan2019a,Ji2016,Mueller2017,Rawlings2017,Allan2020a}.
However, until recently in \cite{Allan2020a}, only conservative stability guarantees could be given for nonlinear MHE, and disturbance gains that increase with increasing $N$ were obtained \cite{Mueller2017,Allan2019a}.
Such a behavior is counter-intuitive and therefore undesirable, since one would naturally expect better estimation results if the respective horizon is enlarged and thus more information is taken into account, which can be also observed in practice.
This gap in the theory could be closed in \cite{Knuefer2018} by choosing a novel cost function for MHE that includes an additional discount factor and thus directly links the cost function to the definition of nonlinear detectability.
Through this direct coupling, a much less restrictive proof technique and thus improved theoretical guarantees became possible, leading to disturbance gains that are uniformly valid for all $N$ and a decay rate that improves with increasing $N$ \cite[Thm. 3]{Knuefer2018}, compare also \cite{Knuefer2021}.
Motivated by \cite{Knuefer2018,Knuefer2021}, we consider a second (time-discounted) cost function throughout this paper that will be specified in detail in the following definition.
As we will show in Sections~\ref{sec:candidate_nominal}-\ref{sec:discussion}, the results from the literature using the standard (non-discounted) cost function \cite{Mueller2017,Allan2019a}, and the time-discounted cost function \cite{Knuefer2021} remain qualitatively valid for our suboptimal setup in terms of the dependence of disturbance gains on horizon length.

\begin{defn}[Time-discounted cost function] \label{def:costfunction_td}
	Let $t \in \mathbb{I}_{\geq 0}$, $N\in \mathbb{I}_{\geq1}$, some prior $\bar{x}_{t-\mathcal{N}}  \in \mathbb{X}$ and the input and output sequences $\boldsymbol{u}_t$ and $\boldsymbol{y}_t$ of system (\ref{eq:system}) in the time interval $\mathbb{I}_{[t-\mathcal{N},t-1]}$ be given and let $\bar{\eta} \in (0,1)$, $\Gamma: \mathbb{X} \times \mathbb{X} \rightarrow \mathbb{R}_{\geq0}$ and $l: \mathbb{W} \times \mathbb{V} \times \mathbb{R}^{n_y} \rightarrow \mathbb{R}_{\geq0}$.
	For $\chi_{t-\mathcal{N}|t} \in \mathbb{X}$, $\boldsymbol{\omega}_t \in \mathbb{W}^\mathcal{N}$ and $\boldsymbol{\nu}_t \in \mathbb{V}^\mathcal{N}$, define
	\begin{align} \label{eq:costfunction_td}
	J_{\text{td}}(\chi_{t-N|t}, \boldsymbol{\omega}_t, \boldsymbol{\nu}_t) &:= 
	\bar{\eta}^\mathcal{N} \Gamma(\chi_{t-\mathcal{N}}, \bar{x}_{t-\mathcal{N}})\\
	& \ \ \ \ + \sum_{i=1}^\mathcal{N}\bar{\eta}^{i} 
	l(\omega_{t-i|t}, \nu_{t-i|t}, y_{t-i} - \zeta_{t-i|t}). \notag
	\end{align}                                                                                                                                                                        
\end{defn}

Analogous to Assumption~\ref{ass:cost}, we impose positive definiteness of the cost function and link it to the e-IOSS condition.

\begin{assum} \label{ass:cost_td}
	The functions $\Gamma$ and $l$ from Definition~\ref{def:costfunction_td} satisfy Assumption~\ref{ass:cost} with $a=1$, and  $\eta\leq\bar{\eta}$ with $\eta$ from (\ref{eq:eIOSS}).
\end{assum}

\begin{rem} \label{rem:implementing}
	Note that due to Assumption~\ref{ass:cost_td}, the cost function~(\ref{eq:costfunction_td}) is in general not differentiable at the origin. 
	The solver of the corresponding nonlinear program (NLP) must therefore be chosen with care, e.g., by using derivative-free optimization methods \cite{Rios2012}.
	{Alternatively, we can exploit the Euclidean norm in~\eqref{eq:costfunction_td} and introduce additional decision variables and appropriate constraints in order to transform this into a larger NLP, but with a cost function and constraints that are differentiable on their respective domains, so that standard gradient-based solvers can still be applied.}
\end{rem}

Now, rather than solving (\ref{eq:optiMHE}) to optimality at each time $t \in \mathbb{I}_{\geq 0}$, we consider the following suboptimal estimator.

\begin{defn}[Suboptimal estimator] \label{def:estimator}
	Let $t \in \mathbb{I}_{\geq 0}$, $N\in \mathbb{I}_{\geq1}$, some prior $\bar{x}_{t-\mathcal{N}}  \in \mathbb{X}$ and the input and output sequences $\boldsymbol{u}_t$ and $\boldsymbol{y}_t$ of system (\ref{eq:system}) in the time interval $\mathbb{I}_{[t-\mathcal{N},t-1]}$ be given and let $(\tilde{x}_{t-\mathcal{N}|t}, \tilde{\boldsymbol{w}}_t,\tilde{\boldsymbol{v}}_t) \in \mathbb{X} \times \mathbb{W}^\mathcal{N} \times \mathbb{V}^\mathcal{N}$ denote a feasible \textit{candidate solution} to the MHE problem (\ref{eq:optiMHE})-(\ref{eq:constraints}).	
	Then, the corresponding suboptimal solution of (\ref{eq:optiMHE}) is defined as \textit{any} $( \hat{x}_{t-\mathcal{N}|t}, \hat{\boldsymbol{w}}_t, \hat{\boldsymbol{v}}_t ) \in \mathbb{X} \times \mathbb{W}^\mathcal{N} \times \mathbb{V}^\mathcal{N}$ that satisfies (i) the MHE constraints (\ref{eq:constraints}) and (ii) the cost decrease condition
	\begin{equation} \label{eq:estimator}
	J(\hat{x}_{t-\mathcal{N}|t}, \hat{\boldsymbol{w}}_t,\hat{\boldsymbol{v}}_t) \leq 
	J(\tilde{x}_{t-\mathcal{N}|t}, \tilde{\boldsymbol{w}}_t, \tilde{\boldsymbol{v}}_t).
	\end{equation}
	The state estimate at time $t \in \mathbb{I}_{\geq 0}$ is defined as
	$\hat{x}_t = \hat{x}_{t|t}$.
\end{defn}                                                                                                                                                                                                                     

\begin{rem}\label{rem:decrease_condition}
	Note that (\ref{eq:estimator}) ensures that at a given time $t$, the cost of a suboptimal solution is no larger than the cost of the candidate solution.
	This can be guaranteed in general by nearly all numerical solvers applied to (\ref{eq:optiMHE}) subject to (\ref{eq:constraints}) and (\ref{eq:estimator}), if they are initialized with the candidate solution as a warm start and then terminated after a finite number of iterations (including 0), cf. \cite{Pannocchia2011a}. 
	To this end, one may implement (\ref{eq:estimator}) as an additional constraint and use some algorithm that provides, at every iteration, a feasible estimate, which is satisfied by, e.g., feasible sequential quadratic programming (fSQP) algorithms, cf. \cite{Lawrence2001}.
	{Alternatively, one could explicitly verify (\ref{eq:constraints}) and (\ref{eq:estimator}) for a given suboptimal solution $(\hat{x}_{t-\mathcal{N}|t}, \hat{\boldsymbol{w}}_t, \hat{\boldsymbol{v}}_t)$ and, if at least one of these conditions is violated, choose the candidate solution as the current suboptimal solution (which satisfies all constraints by definition).}
	{We point out that this does not require warm starting the optimizer with the candidate solution itself; instead, any warm start could be chosen, e.g., based on the shifted solution from one time step before, extended with a one-step forward prediction using the system dynamics, cf.~\cite{Wynn2014,Kuehl2011}.
	Taking such an improved warm start into account while having the candidate solution $(\tilde{x}_{t-\mathcal{N}|t}, \tilde{\boldsymbol{w}}_t,\tilde{\boldsymbol{v}}_t)$ in hand to ensure condition \eqref{eq:estimator} and thus guarantee robustly stable estimation can yield improved performance in practice, compare also the simulation example in Section~\ref{sec:simulation}.}
\end{rem}	

The aim of this work is to show that the suboptimal estimator from Definition~\ref{def:estimator} is robustly stable by means of the following notion of robust global exponential stability (RGES).

\begin{defn}[RGES] \label{def:RGES}
	A state estimator for system~(\ref{eq:system}) is RGES if there exist constants $C_1,C_2,C_3>0$ and $\lambda \in (0,1)$ such that the produced estimate $\hat{x}_t$ at time $t \in \mathbb{I}_{\geq0}$ satisfies
	\begin{equation}\label{eq:RGES} 
	| x_t - \hat{x}_t | \leq
	C_1|x_0 - \hat{x}_0|\lambda^{t} + \sum_{\tau=1}^{t}\lambda^{\tau}
	\big(C_2|w_{t-\tau}| + C_3|v_{t-\tau}|
	\big)
	\end{equation} 
	for all initial conditions $x_0,\hat{x}_0 \in \mathbb{X}$ and all disturbance sequences $\boldsymbol{w} \in \mathbb{W}^{\infty}$ and $\boldsymbol{v} \in \mathbb{V}^{\infty}$.
\end{defn}

This definition of robust stability corresponds to our notion of detectability from (\ref{eq:eIOSS}) in the sense that it includes the discounting of disturbances, and has already been adequately studied in, e.g., \cite{Knuefer2020,Knuefer2021,Allan2021,Allan2020a,Rawlings2017}.
{Note that unlike the asymptotic-gain formulation of robust stability used previously in, e.g., \cite{Allan2019a,Mueller2017,Ji2016}, Definition~\ref{def:RGES} also implies that the estimation error of general estimators in the form of \eqref{eq:observer} converges to zero if the disturbances converge to zero, compare also \cite[Sec. 5.3]{Allan2020a}.}

To establish RGES of the suboptimal estimator from Definition~\ref{def:estimator}, we construct the required candidate solution
by the use of an additional auxiliary nonlinear observer, which is part of the following section.

\subsection{Auxiliary nonlinear observer}

Motivated by \cite{Allan2021}, we define the auxiliary observer in terms of a sequence of maps. 

\begin{assum} \label{ass:aux_obs}
	Let $t \in \mathbb{I}_{\geq 0}$ and the initial condition $z_0 \in \mathbb{X}$ be given.
	For any sequences of inputs $\boldsymbol{u}= \lbrace u_0,\dots,u_{t-1} \rbrace \in \mathbb{U}^t$ and outputs $\boldsymbol{y}= \lbrace y_0,\dots,y_{t-1} \rbrace \in \mathbb{Y}^t$ of system (\ref{eq:system}), there exists a sequence of maps $\Psi_t:\mathbb{X} \times \mathbb{U}^{t} \times \mathbb{Y}^{t}\rightarrow \mathbb{X}$ such that 	
	\begin{equation} \label{eq:observer}
		z_{t}=\Psi_{t}(z_0,\boldsymbol{u},\boldsymbol{y})
	\end{equation}
	is an RGES state estimate of system (\ref{eq:system}),
	i.e., there exist constants $C_p, C_w, C_v>0$ and {$\rho \in [\eta,1)$ with $\eta$ from \eqref{eq:eIOSS}} such that
	\begin{equation} \label{eq:observer_RGES}
		| x_t - z_t | \leq
		C_p|x_0 - z_0|\rho^{t} + \sum_{\tau=1}^{t}\rho^{\tau}
		\big(C_w|w_{t-\tau}| + C_v|v_{t-\tau}|
		\big)
	\end{equation} 
	for all $t \in \mathbb{I}_{\geq 0}$, all $x_0 \in \mathbb{X}$, and all $\boldsymbol{w} \in \mathbb{W}^{\infty}$ and $\boldsymbol{v} \in \mathbb{V}^{\infty}$.
\end{assum}

{Note that for ease of presentation, we impose $\mathbb{X} = \mathbb{R}^{n_x}$ in Assumption~\ref{ass:aux_obs} (cf. the discussion below~\eqref{eq:system}), so that we can trivially guarantee that $z_t \in \mathbb{X}$ for all $t \in \mathbb{I}_{\geq 0}$; this in turn obviously requires a globally convergent behavior in~\eqref{eq:observer_RGES}.
In practice, however, condition \eqref{eq:observer_RGES} can often be shown only locally on some set $\overline{\mathbb{X}}\subset\mathbb{R}^{n_x}$.
This is addressed in Section~\ref{sec:constraints}, where we additionally allow for state constraints in the MHE design, compare also the simulation example in Section~\ref{sec:simulation}.
Note also that system~\eqref{eq:observer} is a very general description of an observer, not even requiring a classical state space representation, cf.~\cite{Allan2021}.
In practice, Assumption~\ref{ass:aux_obs} can be verified using, e.g., nonlinear Luenberger-like observers \cite{Gauthier1992,Zemouche2013,Yi2021}, or Kalman-like observers, e.g., \cite{Reif1999,Jaganath2005}, compare also Remark~\ref{rem:fullorder} below.}

To connect the auxiliary observer to the suboptimal estimator, we suggest the following initialization method.
Similar to the receding horizon fashion of MHE, we define $\mathcal{T}:= \min \lbrace t, T \rbrace$ for some fixed $T \in {\mathbb{I}_{>N}}$ and re-initialize the auxiliary observer, at each time $t \in \mathbb{I}_{\geq 0}$, using the given past suboptimal state estimate $\hat{x}_{t-\mathcal{T}}$ according to
\begin{equation}\label{eq:init}
	z_{t-\mathcal{T}|t} = \hat{x}_{t-\mathcal{T}}.
\end{equation}
{The following lemma provides an upper bound on the observer error in the respective time interval $\mathbb{I}_{[t-\mathcal{T},t-1]}$.}

\begin{lem}\label{cor:obs}
	Suppose Assumption \ref{ass:aux_obs} applies. Let some $T \in {\mathbb{I}_{>N}}$ and a past suboptimal state estimate $\hat{x}_{t-\mathcal{T}}$ be given. 
	If (\ref{eq:init}) holds, then the estimation error of observer~(\ref{eq:observer}) satisfies 
	\begin{align}\label{eq:RGES_observer}
	|x_{t-i}-z_{t-i|t} | &\leq
	C_p|x_{t-\mathcal{T}} - \hat{x}_{t-\mathcal{T}}|\rho^{\mathcal{T}-i} \\
	& \ \ \ \ + \sum_{\tau=i+1}^{\mathcal{T}}\rho^{\tau-i} 
	\big(C_w |w_{t-\tau}| + C_v |v_{t-\tau}|\big) \notag
	\end{align}
	for all $t \in \mathbb{I}_{\geq 0}$, all $i \in \mathbb{I}_{[0,\mathcal{T}]}$, and all $x_0, \hat{x}_0 \in \mathbb{X}$, $\boldsymbol{w} \in \mathbb{W}^{\infty}$, and $\boldsymbol{v} \in \mathbb{V}^{\infty}$.
\end{lem}

\begin{IEEEproof}
	This follows directly from (\ref{eq:observer_RGES}) with respect to the initialization of the observer (\ref{eq:init}).
\end{IEEEproof}

{We use the auxiliary observer~(\ref{eq:observer}) initialized via \eqref{eq:init}} not only to construct the candidate solution to the MHE problem, but also to define the prior in the cost functions (\ref{eq:costfunction}) and (\ref{eq:costfunction_td}) by setting 
\begin{equation} \label{eq:prior}
	\bar{x}_{t-\mathcal{N}} = z_{t-\mathcal{N}|t}, \qquad {t \in \mathbb{I}_{\geq 0}.}
\end{equation}

The overall idea to establish robust stability of the proposed suboptimal estimator is now to choose $T>N$ and to exploit the contraction property  of the auxiliary observer from $t-T$ to $t-N$ through the candidate solution.
We will show that, under certain conditions, there exists some $T$ large enough such that the suboptimal estimator is RGES even in the case of zero iterations solving the corresponding NLP.
To this end, in Section~\ref{sec:candidate_nominal}, we will construct the candidate solution based on the nominal system (\ref{eq:system}) initialized at an estimate provided by the auxiliary observer (\ref{eq:observer}).
Furthermore, assuming one-step controllability with respect to the disturbance $w$ in (\ref{eq:system_a}) enables us to construct a more sophisticated candidate solution that also incorporates the latest estimates of the auxiliary observer.
As we will show in Section~\ref{sec:candidate_observer}, this can both allow for improved theoretical guarantees (in particular, disturbance gains that are uniformly valid for all $N$) and, as will be seen in Section~\ref{sec:simulation}, lead to improved estimation results in practice.

\section{Candidate solution: nominal trajectory}\label{sec:candidate_nominal}

In this section, we construct the required candidate solution based on the nominal system (\ref{eq:system}) initialized with a past estimate obtained from the auxiliary observer (\ref{eq:observer}). 
We will prove RGES of the proposed suboptimal estimator using both the non-discounted cost function (Section~\ref{sec:NT_ND}) and the time discounted cost function (Section~\ref{sec:NT_TD}).
We define
\begin{equation}\label{eq:candidate_NL}
	(\tilde{x}_{t-\mathcal{N}|t},\tilde{\boldsymbol{w}}_t, \tilde{\boldsymbol{v}}_t)
	= ( z_{t-\mathcal{N}|t}, \boldsymbol{0}, \boldsymbol{0} ),                                                                                                                                                                                              
\end{equation}
which yields $(\tilde{\boldsymbol{x}}_t, \boldsymbol{u}_t, \tilde{\boldsymbol{w}}_t, \tilde{\boldsymbol{v}}_t, \tilde{\boldsymbol{y}}_t ) \in \Sigma^{\mathcal{N}}$ under (\ref{eq:system}).
{The steps that need to be performed at each time $t\in\mathbb{I}_{\geq0}$ can be sum\-ma\-rized as follows.}
\begin{algorithm}
	\caption{{Suboptimal MHE}}
	\label{alg:sMHE}
	\begin{algorithmic}[1]
		\State {Collect the current data sequences $\{u_{t-\mathcal{T}},\ldots,u_{t-1}\}$ and $\{y_{t-\mathcal{T}},\ldots,y_{t-1}\}$ and the past suboptimal estimate $\hat{x}_{t-\mathcal{T}}$.}
		\State {Re-initialize the auxiliary observer via \eqref{eq:init} and perform a forward simulation for $\mathcal{T}-\mathcal{N}$ steps via \eqref{eq:observer}.}
		\State {Construct the candidate solution \eqref{eq:candidate_NL}.}
		\State {Approximately solve MHE problem \eqref{eq:optiMHE} with prior \eqref{eq:prior} and cost function \eqref{eq:costfunction} (or \eqref{eq:costfunction_td}) subject to \eqref{eq:constraints} and \eqref{eq:estimator}.}
		\State {Obtain new suboptimal estimate: $\hat{x}_t = \hat{x}_{t|t}$.}
		\State {Set $t=t+1$ and go back to 1.}
	\end{algorithmic}
\end{algorithm}

For the stability analysis, we need an additional continuity assumption on the function $f$ as stated in the following.

\begin{assum}\label{ass:lipschitz_f}
	The function $f$ is Lipschitz continuous, i.e., there exists some constant $F > 0$ such that $|f(x,u,w)-f(\chi,\mu,\omega)|\leq F(|x-\chi|+|u-\mu|+|w-\omega|)$ for all $x,\chi \in \mathbb{X}$, $u,\mu \in \mathbb{U}$, and $w,\omega \in \mathbb{W}$.	
\end{assum} 

Before we show robust stability of the proposed estimator, we provide a result that suitably bounds the fitting error of the candidate solution.
Throughout the remainder of this paper, we assume without loss of generality that $F\geq1$, which allows for simpler proofs.
Note that this assumption can also be omitted at the expense of additional case distinctions in the proof of Lemma~\ref{lem:boundedness_of_v} in order to obtain less conservative results.

\begin{lem} \label{lem:boundedness_of_v}
	Suppose that Assumptions \ref{ass:lipschitz}, \ref{ass:aux_obs}, and \ref{ass:lipschitz_f} apply. Let $N\in \mathbb{I}_{\geq 1}$ and $T\in \mathbb{I}_{> N}$ be arbitrary.
	Then, the fitting error of the trajectory defined by the candidate solution (\ref{eq:candidate_NL}) satisfies
	\begin{align} \label{eq:leF_wv}
	& |y_{t-i}-\tilde{y}_{t-i|t}| 
	\leq \sigma_{\mathcal{N}} F^{-i}
	\Big(
	C_p|x_{t-\mathcal{T}}-\hat{x}_{t-\mathcal{T}}|\rho^{\mathcal{T}} \\
	& \ \ \ 
	+  \sum_{\tau=i}^{\mathcal{T}}\rho^{\tau}\big(C_w|w_{t-\tau}| 
	+ C_v |v_{t-\tau}|\big) 
	\Big) \quad \forall i \in \mathbb{I}_{[1,\mathcal{N}]} \notag
	\end{align} 
	for all $t \in \mathbb{I}_{\geq 0}$, with $\sigma_{\mathcal{N}} =HC (F/\rho)^{\mathcal{N}}$ and $C:=\max \lbrace 1, C_w^{-2}, C_v^{-2} \rbrace$.
\end{lem}

\begin{IEEEproof}
	Since the candidate solution defines a trajectory of system (\ref{eq:system}), we can apply
	the output equation (\ref{eq:system_b}).
	Together with Assumption \ref{ass:lipschitz}, the fitting error can be bounded by
	\begin{equation} 
	|y_{t-i}-\tilde{y}_{t-i|t}| \leq H\left( |x_{t-i} - \tilde{x}_{t-i|t}| + |v_{t-i}| \right) \label{eq:leF_wv_1}
	\end{equation}
	for all $t \in \mathbb{I}_{\geq 0}$ and all $i \in \mathbb{I}_{[1,\mathcal{N}]}$.
	Applying (\ref{eq:system_a}) together with Assumption~\ref{ass:lipschitz_f}, by induction we can show that
	\begin{equation} \label{eq:leF_wv_2}
	|x_{t-i} - \tilde{x}_{t-i|t}| \leq  F^{\mathcal{N}-i}|x_{t-\mathcal{N}} - \tilde{x}_{t-\mathcal{N}|t}| + \sum_{\tau = i+1}^{\mathcal{N}}F^{\tau-i}|w_{t-\tau}|.
	\end{equation}
	Since $\tilde{x}_{t-\mathcal{N}|t} = z_{t-\mathcal{N}|t}$ due to (\ref{eq:candidate_NL}), we can exploit {Lemma}~\ref{cor:obs} and apply (\ref{eq:RGES_observer}) evaluated at time $i = \mathcal{N}$.
	Using the definition of $C$ as stated in this Lemma and the fact that $F\geq1$ then yields
	\begin{align}
	|x_{t-i} - \tilde{x}_{t-i|t}|
	&\leq \sqrt{C}F^{\mathcal{N}-i}\rho^{-\mathcal{N}} 
	\Big(
	C_p|x_{t-\mathcal{T}} - z_{t-\mathcal{T}|t}|\rho^{\mathcal{T}}  \notag \\
	&\ \ \ + \sum_{\tau=i+1}^{\mathcal{T}}\rho^{\tau}
	\big(C_w |w_{t-\tau}| + C_v |v_{t-\tau}| \big)
	\Big). \label{eq:leF_wv_3}
	\end{align}
	By applying (\ref{eq:leF_wv_3}) to (\ref{eq:leF_wv_1}) and due to the initialization of the auxiliary observer (\ref{eq:init}), we obtain
	\begin{align*} 
	|y_{t-i}-\tilde{y}_{t-i|t}| \notag
	& \leq HCF^{\mathcal{N}-i}\rho^{-\mathcal{N}} 
	\Big(
	C_p|x_{t-\mathcal{T}}-\hat{x}_{t-\mathcal{T}}|\rho^{\mathcal{T}} \\
	& \ \ + \sum_{\tau=i}^{\mathcal{T}}\rho^{\tau}
	\big(C_w |w_{t-\tau}| + C_v |v_{t-\tau}| \big)
	\Big).
	\end{align*}
	Defining $\sigma_{\mathcal{N}}$ as stated in this Lemma yields (\ref{eq:leF_wv}), which finishes this proof.
\end{IEEEproof}

The following sections provide stability guarantees for the suboptimal estimator from Definition \ref{def:estimator} using both the non-discounted cost function (\ref{eq:costfunction}) and the time-discounted cost function (\ref{eq:costfunction_td}) together with the candidate solution (\ref{eq:candidate_NL}).

\subsection{Non-discounted cost function}\label{sec:NT_ND}
We first consider the non-discounted cost function (\ref{eq:costfunction}).
Before we can state the desired result, we first need two auxiliary lemmas that provide a bound on the cost function and on the estimation error both evaluated at a given estimated suboptimal trajectory.

\begin{lem} \label{lem:boundedness_of_J}
	Suppose that Assumptions \ref{ass:lipschitz}, \ref{ass:cost}, \ref{ass:aux_obs}, and \ref{ass:lipschitz_f} apply. Let $N\in \mathbb{I}_{\geq 1}$ and $T\in \mathbb{I}_{> N}$ be arbitrary.
	Then, there exists some $\bar{\sigma}_{\mathcal{N}}>0$ such that the {cost function \eqref{eq:costfunction}} evaluated at any suboptimal estimate provided by the estimator from Definition~\ref{def:estimator} using the candidate solution (\ref{eq:candidate_NL}) satisfies 
	\begin{align} \label{eq:lem_J}
	J_{\text{nd}}(\hat{x}_{t-\mathcal{N}|t}, \hat{\boldsymbol{w}}_t, \hat{\boldsymbol{v}}_t) &\leq 
	\bar{\sigma}_{\mathcal{N}}
	\Big(C_p|x_{t-\mathcal{T}} - \hat{x}_{t-\mathcal{T}}|\rho^{\mathcal{T}} \\
	& \ \ \ \ + 
	\sum_{\tau=1}^{\mathcal{T}}\rho^{\tau}
	\big(C_w|w_{t-\tau}| + C_v|v_{t-\tau}|\big)
	\Big)^a \notag
	\end{align}
	for all $t \in \mathbb{I}_{\geq 0}$.
\end{lem}

\begin{IEEEproof}
	We start from (\ref{eq:estimator}). By (\ref{eq:cost}) and our choices of candidate solution (\ref{eq:candidate_NL}) and prior (\ref{eq:prior}), we obtain
	\begin{equation*}
	J_{\text{nd}}(\hat{x}_{t-\mathcal{N}|t}, \hat{\boldsymbol{w}}_t, \hat{\boldsymbol{v}}_t )
	\leq
	\sum_{i = 1}^{\mathcal{N}}\overline{c}_y|y_{t-i}-\tilde{y}_{t-i|t}|^a
	\end{equation*}
	for all $t \in \mathbb{I}_{\geq0}$.
	Applying Lemma~\ref{lem:boundedness_of_v} yields
	\begin{align}
	J_{\text{nd}}(&\hat{x}_{t-\mathcal{N}|t}, \hat{\boldsymbol{w}}_t, \hat{\boldsymbol{v}}_t) 
	\leq \overline{c}_y
	\sigma_{\mathcal{N}}^a \sum_{i = 1}^{\mathcal{N}} F^{-ai}
	\Big(
	C_p|x_{t-\mathcal{T}}-\hat{x}_{t-\mathcal{T}}|\rho^{\mathcal{T}} \notag  \\
	& \qquad +  \sum_{\tau=i}^{\mathcal{T}}\rho^{\tau}\big(C_w|w_{t-\tau|t}| 
	+ C_v|v_{t-\tau|t}|\big)
	\Big)^a. \label{eq:lem_J_2}
	\end{align}
	Note that the argument of the inner sum of the double sum in~(\ref{eq:lem_J_2}) is independent of $i$, and hence we can enlarge the lower bound of summation to $\tau=1$ and move the complete term in large brackets to the power of $a$ in front of the outer sum.
	Then, by the geometric series we note that
	\begin{equation*}
	\sum_{i = 1}^{\mathcal{N}}F^{-ai} 
	= \frac{F^{-a}-F^{-a(\mathcal{N}-1)}}{1-F^{-a}} = \frac{1-F^{-a\mathcal{N}}}{F^a-1}
	\end{equation*}
	for $F\neq 1$.
	Applying the definition
	\begin{equation}\label{eq:barsigmaN}
	\bar{\sigma}_{\mathcal{N}} := \overline{c}_y\sigma_{\mathcal{N}}^a \times \begin{cases}
	\frac{1-F^{-a\mathcal{N}}}{F^a-1}, & F \neq 1 \\
	\mathcal{N}, & F = 1			
	\end{cases}	
	\end{equation}	
	to (\ref{eq:lem_J_2}) yields (\ref{eq:lem_J}), which completes this proof.	
\end{IEEEproof}

\begin{lem} \label{lem:boundedness_of_e}
	Suppose that system (\ref{eq:system}) is e-IOSS and that Assumptions \ref{ass:lipschitz}, \ref{ass:cost}, \ref{ass:aux_obs}, and \ref{ass:lipschitz_f} apply. Let $N\in \mathbb{I}_{\geq 1}$ and $T\in \mathbb{I}_{> N}$ be arbitrary.
	Then, there exist $C_{\mathcal{N},1}, C_{\mathcal{N},2}, C_{\mathcal{N},3} > 0$ such that the estimation error of the estimator from Definition~\ref{def:estimator} using the {cost function~\eqref{eq:costfunction}} and the candidate solution (\ref{eq:candidate_NL}) satisfies
	\begin{align}  \label{eq:lem_e}
	|x_{t}-\hat{x}_{t}| &\leq
	C_{\mathcal{N},1}|x_{t-\mathcal{T}}-\hat{x}_{t-\mathcal{T}}|\rho^{\mathcal{T}} \\
	& \ \ \ \ \ +  \sum_{\tau=1}^{\mathcal{T}} \rho^{\tau}
	\big(C_{\mathcal{N},2}|w_{t-\tau}|	+ C_{\mathcal{N},3}|v_{t-\tau}|\big) \notag
	\end{align}
	for all $t \in \mathbb{I}_{\geq 0}$.
\end{lem}

\begin{IEEEproof}
	Since both the real and the estimated trajectory are trajectories of system (\ref{eq:system}), we can describe the deviation of their states at any $t \in \mathbb{I}_{\geq 0}$ starting at time $t-\mathcal{N}$ by utilizing the e-IOSS condition (\ref{eq:eIOSS}). More precisely, consider (\ref{eq:eIOSS}), the real trajectory starting at $x_{t-\mathcal{N}}$ driven by the sequences $\boldsymbol{u}_{t}$, $\boldsymbol{w}_{t}$ and $\boldsymbol{v}_{t}$, and the estimated (suboptimal) trajectory starting at $\hat{x}_{t-\mathcal{N}|t}$ driven by the sequences $\boldsymbol{u}_{t}$, $\hat{\boldsymbol{w}}_{t}$ and $\hat{\boldsymbol{v}}_{t}$.
	By applying the triangle inequality, it further follows that	
	\begin{align}
		&|x_t - \hat{x}_t| 
		\leq  c_p\eta^\mathcal{N}|\hat{x}_{t-\mathcal{N}|t} - \bar{x}_{t-\mathcal{N}}| \label{eq:lem_e_1} \\
		& \ \ \ \ 
		+ c_p\eta^\mathcal{N}|x_{t-\mathcal{N}} - \bar{x}_{t-\mathcal{N}}| 
		+ \sum_{\tau=1}^\mathcal{N}\eta^{\tau} 
		\big(c_w|w_{t-\tau}| + c_v|v_{t-\tau}| \big) \notag \\ 
		& \ \ \ \ 
		+ \sum_{\tau=1}^\mathcal{N}\eta^{\tau} 
		\big( c_w|\hat{w}_{t-\tau|t}| + c_v|\hat{v}_{t-\tau|t}| + c_y|y_{t-\tau}-\hat{y}_{t-\tau|t}|\big) \notag	
	\end{align}
	for all $t \in \mathbb{I}_{\geq 0}$.
	Now the objective is to find suitable upper bounds for the estimates in (\ref{eq:lem_e_1}).
	By the choice of the prior in (\ref{eq:prior}), for the second term of the right-hand side in (\ref{eq:lem_e_1}) we can apply (\ref{eq:RGES_observer}) evaluated at time $i = \mathcal{N}$, which yields
	\begin{align}
	|x_{t-\mathcal{N}} - \bar{x}_{t-\mathcal{N}}|
	&\leq
	\rho^{-\mathcal{N}}\Big(
	C_p|x_{t-\mathcal{T}} - \hat{x}_{t-\mathcal{T}}|\rho^{\mathcal{T}} \label{eq:lem_e_4} \\
	& \ \ \ \ \ + \sum_{\tau=\mathcal{N}+1}^{\mathcal{T}}\rho^{\tau}\big(C_w|w_{t-\tau}| 
	+ C_v|v_{t-\tau}|\big)
	\Big). \notag
	\end{align}
	For the remaining terms on the right-hand side in (\ref{eq:lem_e_1}), we consider the lower bound on the suboptimal cost given by Assumption~\ref{ass:cost}. By applying (\ref{eq:cost})-(\ref{eq:cost_lb}) and using the fact that $\eta<1$, we obtain
	\begin{align*} 
	&J_{\text{nd}}(\hat{x}_{t-\mathcal{N}|t}, \hat{\boldsymbol{w}}_t, \hat{\boldsymbol{v}}_t) 
	\geq	
	c_p\eta^\mathcal{N}|\hat{x}_{t-\mathcal{N}|t} - \bar{x}_{t-\mathcal{N}}|^a  \\
	& \  + \sum_{i=1}^\mathcal{N}\eta^i\big(c_w|\hat{w}_{t-i|t}|^a                               
	+ c_v|\hat{v}_{t-i|t}|^a+ c_y|y_{t-i}-\hat{y}_{t-i|t}|^a \big).\\
	&\geq                                                             
	\Big(
	c_p\eta^\mathcal{N}+\sum_{i=1}^\mathcal{N}\eta^i\big(c_w+c_v+c_y \big)                                                                                           
	\Big)^{1-a} \\
	& \ \ \ \times \Big( c_p\eta^\mathcal{N}|\hat{x}_{t-\mathcal{N}|t} - \bar{x}_{t-\mathcal{N}}| \\
	& \ \ \ \ \ \  + \sum_{i=1}^\mathcal{N}\eta^i\big(c_w|\hat{w}_{t-i|t}|
	+ c_v|\hat{v}_{t-i|t}|+ c_y|y_{t-i}-\hat{y}_{t-i|t}| \big)\Big)^a,
	\end{align*}
	where the last step follows from\footnote{Note that $r \rightarrow r^a$ is convex for $r\geq0$ since $a\geq1$ by Assumption~\ref{ass:cost}.} Jensen's inequality.
	Now raising both sides to the power of $\alpha:=1/a$ and applying the geometric series yields
	\begin{align}
	&( c_\mathcal{N}J_{\text{nd}}(\hat{x}_{t-N|t}, \hat{\boldsymbol{w}}_t))^{\alpha} \label{eq:lem_e_5} 
	\geq
	c_p\eta^{\mathcal{N}}|\hat{x}_{t-\mathcal{N}|t} - \bar{x}_{t-\mathcal{N}}| \\
	& \ \ + \sum_{i=1}^\mathcal{N}\eta^i\big(c_w|\hat{w}_{t-i|t}| + c_v|\hat{v}_{t-i|t}| + c_y|y_{t-i}-\hat{y}_{t-i|t}|\big) \notag	
	\end{align}
	with
	$c_\mathcal{N} := \big(c_p\eta^\mathcal{N} + (\eta-\eta^{\mathcal{N}+1})(1-\eta)^{-1}(c_w + c_v + c_y)\big)^{a-1}$. 
	Note that the right-hand side of (\ref{eq:lem_e_5}) corresponds to the terms in the first and third line of the right-hand side of (\ref{eq:lem_e_1}).
	Now we exploit Lemma~\ref{lem:boundedness_of_J} to upper bound the suboptimal cost in (\ref{eq:lem_e_5}) and apply the result together with (\ref{eq:lem_e_4}) to (\ref{eq:lem_e_1}), and we obtain (\ref{eq:lem_e}) with
	\begin{subequations}
		\label{eq:gains_N}
		\begin{align} 
		C_{\mathcal{N},1} &:=
		(c_p(\eta/\rho)^\mathcal{N} 
		+ (c_\mathcal{N} \bar{\sigma}_{\mathcal{N}})^{\alpha})C_p,\\
		C_{\mathcal{N},2} &:=
		(c_p(\eta/\rho)^\mathcal{N} 	
		+ (c_\mathcal{N} \bar{\sigma}_{\mathcal{N}})^{\alpha})C_w
		+ c_w\eta/\rho,\\
		C_{\mathcal{N},3} &:=
		(c_p(\eta/\rho)^\mathcal{N} 
		+ (c_\mathcal{N} \bar{\sigma}_{\mathcal{N}})^{\alpha})C_v + c_v\eta/\rho,
		\end{align}
	\end{subequations}
	which concludes this proof.
\end{IEEEproof}

Now we are in a position to state our first main result.

\begin{thm}\label{theorem:NL_1}
	Suppose that system (\ref{eq:system}) is e-IOSS and that Assumptions \ref{ass:lipschitz}, \ref{ass:cost}, \ref{ass:aux_obs}, and \ref{ass:lipschitz_f} apply. Choose $N\in \mathbb{I}_{\geq 1}$ arbitrarily and $T \in \mathbb{I}_{> N}$ such that
	\begin{equation} \label{eq:def_C1}
	T > -\frac{\ln C_1}{\ln\rho}, \qquad C_1 = \max_{\mathcal{N} \in \mathbb{I}_{[0,N]}}C_{\mathcal{N},1}. 
	\end{equation}
	Then, the suboptimal moving horizon estimator from Definition~\ref{def:estimator} using the cost function from Definition~\ref{def:costfunction} and the candidate solution (\ref{eq:candidate_NL}) is RGES and satisfies (\ref{eq:RGES}) for all $t \in \mathbb{I}_{\geq0}$, where the decay rate is given by $\lambda=\sqrt[T]{C_1}\rho$.
\end{thm}

\begin{IEEEproof}
	From Lemma~\ref{lem:boundedness_of_e} and the definition of $C_1$, $T$, and $\lambda$, it follows that 
	\begin{align*}  
		|x_{t+T} - \hat{x}_{t+T}| &\leq
		|x_{t} - \hat{x}_{t}|\lambda^{T} \\
		& \ \ \ + \sum_{\tau=1}^{T} \lambda^{\tau}
		\big(C_2|w_{t+T-\tau}|+ C_3|v_{t+T-\tau}|\big)
	\end{align*}
	for $t \in \mathbb{I}_{\geq 0}$ {with $C_i := C_1^{-1/T} \max_{\mathcal{N} \in \mathbb{I}_{[0,N]}}C_{\mathcal{N},i}, i \in \lbrace 2, 3 \rbrace$}.
	{By induction and Lemma \ref{lem:boundedness_of_e}, one can show that}
	\begin{align*}  
	|x_{t+kT} - \hat{x}_{t+kT}| & \leq
	C_1|x_{0} - \hat{x}_{0}|\lambda^{t+kT} \\
	& \ \ + \sum_{\tau=1}^{t+kT}\lambda^{\tau}
	\big(C_2|w_{t+kT-\tau}|+C_3|v_{t+kT-\tau}|\big) \notag
	\end{align*}
	for all $t\in \mathbb{I}_{[0,T-1]}$ and all $k\in \mathbb{I}_{\geq 0}$, which is equivalent to~\eqref{eq:RGES} and hence concludes this proof.
\end{IEEEproof}

\begin{rem}
	{By Theorem~\ref{theorem:NL_1}, the proposed suboptimal estimator is RGES for any $N \in \mathbb{I}_{\geq 1}$;} in other words, there is no minimum required horizon length as it is the case in , e.g., \cite{Mueller2017,Knuefer2018,Rawlings2017,Allan2019a,Knuefer2021,Hu2021,Allan2020a}.
	This is due to the fact that we do not require contraction of the estimation error from time $t-N$ to $t$, but establish stability by exploiting the contraction property of the auxiliary observer from time $t-T$ to $t-N$.
\end{rem}

\begin{rem}\label{rem:choice_T}
	{Note that Step 2 in Algorithm~\ref{alg:sMHE}, i.e., the forward simulation of the auxiliary observer for $\mathcal{T}-\mathcal{N}$ steps, is required to compute $z_{t-\mathcal{N}|t}$ and thus both the prior (\ref{eq:prior}) and the candidate solution (\ref{eq:candidate_NL}).}
	To save computation time, however, it is also possible to initialize the auxiliary observer only once at time $t=0$, thus avoiding its repeated re-initialization {and hence Step~2 in Algorithm~\ref{alg:sMHE}}.
	This is a special case of the proposed MHE scheme with $T=t$ and was also considered in the preliminary conference version \cite{Schiller2021}.
	The corresponding estimation error can be bounded by (\ref{eq:lem_e}), and the definitions of $C_i = \max_{\mathcal{N} \in \mathbb{I}_{[0,N]}} C_{\mathcal{N},i}, i \in \lbrace 1, 2, 3 \rbrace$ reveal that, not very surprisingly, suboptimal MHE is RGES for $T=t$. 
	We point out that the decay rate of the estimation error then takes the theoretically best possible value, which is given by $\lambda = \rho$.	
	In contrast, choosing $T$ in (\ref{eq:def_C1}) small results in a worse decay rate $\lambda$ and a slightly more computationally intensive scheme.
	In practice, however, much better performance can be expected since improved suboptimal estimates are used to re-initialize the auxiliary observer, thus introducing additional feedback into the suboptimal estimator.
	This may lead to much faster recovery from a poor initial guess, as also illustrated by the simulation example in Section~\ref{sec:simulation}.
\end{rem}

\subsection{Time-discounted cost function}\label{sec:NT_TD}
We now consider the time-discounted cost function (\ref{eq:costfunction}).
As outlined above Definition~\ref{def:costfunction_td}, this allows for the adoption of a less conservative proof technique, which, as shown below and discussed in more detail in Section~\ref{sec:discussion}, leads to a less restrictive bound on the corresponding estimation error.
The proof of the following theorem consists of three parts: we show that (i) Lemma~\ref{lem:boundedness_of_J} and (ii) Lemma~\ref{lem:boundedness_of_e} still hold for this modified setting, which (iii) allows to proceed as in the proof of Theorem~\ref{theorem:NL_1}.
Note that a similar procedure will also be used in the following section when considering a different candidate solution.

\begin{thm}\label{theorem:NL_2}
	Suppose that system (\ref{eq:system}) is e-IOSS and that Assumptions \ref{ass:lipschitz}, \ref{ass:cost_td}, \ref{ass:aux_obs}, and \ref{ass:lipschitz_f} apply. Choose $N\in \mathbb{I}_{\geq 1}$ arbitrarily.
	Then, there exists $T \in \mathbb{I}_{>N}$ large enough such that the suboptimal moving horizon estimator from Definition~\ref{def:estimator} using the {cost function~\eqref{eq:costfunction_td}} and the candidate solution (\ref{eq:candidate_NL}) is RGES.
\end{thm}

\begin{IEEEproof}
	\textit{Part I.}
	We again start from (\ref{eq:estimator}) with respect to the cost function (\ref{eq:costfunction_td}) and the candidate solution (\ref{eq:candidate_NL}).
	Exploiting Assumption~\ref{ass:cost_td} and our choice of prior (\ref{eq:prior}) then yields
	\begin{equation}\label{eq:lem_J_0}
	J_{\text{td}}(\hat{x}_{t-\mathcal{N}|t}, \hat{\boldsymbol{w}}_t, \hat{\boldsymbol{v}}_t) 
	\leq \sum_{i=1}^{\mathcal{N}}\bar{\eta}^{i}\bar{c}_y|y_{t-i}-\tilde{y}_{t-i|t}| 
	\end{equation}
	for all $t \in \mathbb{I}_{\geq0}$.
	Applying Lemma \ref{lem:boundedness_of_v} leads to
	\begin{align*}
	J_{\text{td}}(\hat{x}_{t-\mathcal{N}|t}, \hat{\boldsymbol{w}}_t, \hat{\boldsymbol{v}}_t)
	&\leq \bar{c}_y\sigma_{\mathcal{N}}
	\sum_{i = 1}^{\mathcal{N}}\bar{\eta}^i F^{-i} \Big( C_p|x_{t-\mathcal{T}}-\hat{x}_{t-\mathcal{T}}|\rho^{\mathcal{T}}  \\
	& \ \ \  + \sum_{\tau=i}^\mathcal{T} \rho^{\tau}\big(C_w|w_{t-\tau}| + C_v|v_{t-\tau}|\big) \Big).
	\end{align*}
	Proceeding as in the proof of Lemma~\ref{lem:boundedness_of_J}, i.e., enlarging the latter sum by changing the lower bound of summation to $\tau=1$, we obtain (\ref{eq:lem_J}) for all $t \in \mathbb{I}_{\geq0}$ with $J_{\text{nd}}$ replaced by $J_{\text{td}}$,  $a = 1$, and where
	$\bar{\sigma}_{\mathcal{N}} 
	= \bar{c}_y\sigma_{\mathcal{N}}
	(1- (\bar{\eta}/F)^{\mathcal{N}})/((F/\bar{\eta})-1)$.\\
	\textit{Part II.} From (\ref{eq:lem_e_1}) and the specific structure of the time-discounted cost function (\ref{eq:costfunction_td}) satisfying Assumption~\ref{ass:cost_td}, it follows that
	\begin{align*}
	&|x_t - \hat{x}_t| 
	\leq  c_p\eta^\mathcal{N} |x_{t-\mathcal{N}} - \bar{x}_{t-\mathcal{N}}| \\
	& \ \ \ \ \ \ + \sum_{\tau=1}^\mathcal{N}\eta^{\tau} 
	\big(c_w|w_{t-\tau}| + c_v|v_{t-\tau}|\big)
	+ J_{\text{td}}(\hat{x}_{t-\mathcal{N}|t}, \hat{\boldsymbol{w}}_t, \hat{\boldsymbol{v}}_t)	
	\end{align*}
	for all $t \in \mathbb{I}_{\geq0}$.
	Since (\ref{eq:prior}) holds, we can again apply (\ref{eq:RGES_observer}) evaluated at time $i=t-\mathcal{N}$, which is given by (\ref{eq:lem_e_4}).
	Exploiting the result from the first part of this proof yields (\ref{eq:lem_e}) with 
	\begin{subequations}
		\label{eq:gains_d1}
		\begin{align} 
		C_{\mathcal{N},1} &= (c_p(\eta/\rho)^\mathcal{N} + \bar{\sigma}_{\mathcal{N}}) C_p,\\
		C_{\mathcal{N},2} &= (c_p(\eta/\rho)^\mathcal{N} + \bar{\sigma}_{\mathcal{N}}) C_w + c_w\eta/\rho,\\
		C_{\mathcal{N},3} &= (c_p(\eta/\rho)^\mathcal{N} + \bar{\sigma}_{\mathcal{N}}) C_v + c_v\eta/\rho.
		\end{align}	
	\end{subequations}
	\textit{Part III.} By choosing $C_1$, $T$, and $\lambda$ as stated in Theorem~\ref{theorem:NL_1}, we can follow the same steps as in the proof of Theorem~\ref{theorem:NL_1}, which yields the desired result. 
\end{IEEEproof}

\section{Candidate solution: observer trajectory}\label{sec:candidate_observer}

We now construct a second candidate solution based on the entire trajectory of the auxiliary observer within the estimation horizon, and therefore including the most recent observer estimates. 
This more sophisticated approach allows us to avoid many conservative arguments applied in the proof of Lemma~\ref{lem:boundedness_of_v}, which, as we will show below and discuss in more detail in Section~\ref{sec:discussion}, can lead to improved theoretical results.
To this end, we have to strengthen the conditions on the considered class of nonlinear systems and auxiliary observers.
In particular, to be able to reconstruct the exact trajectory given by the auxiliary observer through the candidate solution, we first require one-step controllability with respect to the disturbance $w$ in (\ref{eq:system_a}), cf. \cite[Rem. 2]{Koehler2021}, and second, an auxiliary observer given in output injection form, cf. \cite{Knuefer2020,Sontag1997}.
Therefore, the following assumptions are required to hold. 

\begin{assum}\label{ass:add_dist}
	The perturbed system dynamics (\ref{eq:system_a}) satisfies $f(x,u,w) = f_n(x,u) + w$ with $\mathbb{W} = \mathbb{R}^{n_x}$. 
\end{assum}

\begin{assum}\label{ass:fullorder}
	The observer dynamics (\ref{eq:observer}) satisfies $z_{t+1} = f_n(z_t,u_t) + L(z_t,y_t-h_n(z_t,u_t))$ with the output injection law $L : \mathbb{X} \times \mathbb{Y} \rightarrow \mathbb{X}$, where $L(\cdot,0) = 0$. Moreover, the injection law can be uniformly linearly bounded by $L(z_t,y_t-h_n(z_t,u_t))\leq \kappa|y_t-h_n(z_t,u_t)|$ for some fixed constant $\kappa>0$.	
\end{assum}

\begin{rem}\label{rem:fullorder}
	As\-sump\-tion~\ref{ass:fullorder} consists of two parts.
	First, it requires that the auxiliary observer is a full-order state observer in output injection form, cf.~\cite{Knuefer2020,Sontag1997}. 
	Note that assuming output injection form is not restrictive, since from \cite[Lem.~2]{Knuefer2020}, \cite[Lem.~21]{Sontag1997}, it follows that any robustly stable full-order state observer must in fact have this form.
	The second part states a linear bound on the injection law $L$ depending on the current fitting error of the observer.
	Although this linear bound can be restrictive, we note that this is satisfied by following any observer design that utilizes the injection law $L(z_t,y_t-h_n(z_t,u_t)) = K\cdot(y_t-h_n(z_t,u_t))$, where $K \in\mathbb{R}^{n_x\times n_y}$ is a constant or time-varying matrix that can be suitably bounded.
	{This is the case for nonlinear Luenberger-like observers by following, e.g., a high-gain design (cf. \cite{Gauthier1994}), by treating the nonlinearity as suitably bounded uncertainty and solving a sufficient linear matrix inequality (LMI) condition (cf.~\cite{Zemouche2013}), or by using a differential framework and convex optimization (cf.~\cite{Yi2021}).
	In addition, Assumption~\ref{ass:fullorder} can be (locally) satisfied for the extended Kalman filter under uniform observability and boundedness conditions, cf.~\cite{Reif1999}.}
\end{rem}

In the following, we abbreviate $L(z_{t-i|t},y_t-h_n(z_{t-i|t},u_t))$ by $L_{t-i|t}$ for all $i \in \mathbb{I}_{[1,\mathcal{N}]}$ and $t \in \mathbb{I}_{\geq 0}$.
{Provided that Assumptions~\ref{ass:add_dist} and \ref{ass:fullorder} hold, the system and observer dynamics share the same structure, which allows us to interpret the terms $f_n(z,u)$ and $L$ in the observer dynamics directly as estimates of the terms $f_n(x,u)$ and $w$ in the system dynamics.
We hence choose the candidate solution}
\begin{equation}\label{eq:candidate}
	(\tilde{x}_{t-\mathcal{N}|t}, \tilde{\boldsymbol{w}}_t, \tilde{\boldsymbol{v}}_t)
	= (z_{t-\mathcal{N}|t}, \lbrace L_{t-\mathcal{N}|t}, ... , L_{t-1|t} \rbrace, \boldsymbol{0})
\end{equation}
for all $t \in \mathbb{I}_{\geq 0}$.
{To compute the current state estimate $\hat{x}_t$ at any $t\in\mathbb{I}_{\geq0}$, the steps 2 and 3 from Algorithm~\ref{alg:sMHE} have to be modified according to the following.}
\begin{algorithm}
	\begin{algorithmic}[1]
		\setcounter{ALG@line}{1}
		\State {Re-initialize the auxiliary observer via \eqref{eq:init} and perform a forward simulation for $\mathcal{T}$ steps using the observer dynamics according to Assumption~\ref{ass:fullorder}.}
		\State {Construct the candidate solution \eqref{eq:candidate}.}
	\end{algorithmic}
\end{algorithm}

{To show robust stability of the resulting suboptimal estimator, we again start by bounding the fitting error of the candidate solution, i.e., the trajectory defined by $(\tilde{\boldsymbol{x}}_t, \boldsymbol{u}_t, \tilde{\boldsymbol{w}}_t, \tilde{\boldsymbol{v}}_t, \tilde{\boldsymbol{y}}_t ) \in \Sigma^{\mathcal{N}}$ under (\ref{eq:system}).}

\begin{lem} \label{lem:boundedness_of_V}
	Suppose that Assumptions \ref{ass:lipschitz}, \ref{ass:aux_obs}, \ref{ass:add_dist}, and \ref{ass:fullorder} apply. Let $N\in \mathbb{I}_{\geq 1}$ and $T\in \mathbb{I}_{> N}$ be arbitrary.
	Then, the fitting error of the trajectory defined by the candidate solution (\ref{eq:candidate}) satisfies
	\begin{align} \label{eq:leF_wV1}
	&|y_{t-i}-\tilde{y}_{t-i|t}| 
	\leq {H}c\rho^{-i} \Big( C_p|x_{t-\mathcal{T}} - \hat{x}_{t-\mathcal{T}}|\rho^{\mathcal{T}} \\
	& \ \ \ \ \ + \sum_{\tau=i}^\mathcal{T} \rho^{\tau}\big(C_w|w_{t-\tau}| + C_v|v_{t-\tau}|\big) \Big) \quad \forall i \in \mathbb{I}_{[1,\mathcal{N}]} \notag 
	\end{align} 
	for all $t \in \mathbb{I}_{\geq0}$ with $c=\max \lbrace 1, 1/C_v \rbrace$.
\end{lem}

\begin{IEEEproof}
	Due to Assumptions~\ref{ass:add_dist} and \ref{ass:fullorder} and the candidate solution (\ref{eq:candidate}), we have that $\tilde{x}_{t-i|t} = z_{t-i|t}$ for all $t \in \mathbb{I}_{\geq0}$ and all $i \in \mathbb{I}_{[1,\mathcal{N}]}$. 
	Hence, from Assumption~\ref{ass:lipschitz} and using $\tilde{v}_t = 0$ from (\ref{eq:candidate}), we obtain $|y_{t-i}-\tilde{y}_{t-i|t}| \leq H(|x_{t-i} - z_{t-i|t}| + |v_{t-i}|)$. 
	Since the auxiliary observer is RGES by Assumption~\ref{ass:aux_obs}, we can apply (\ref{eq:RGES_observer}) for all $i \in \mathbb{I}_{[1,\mathcal{N}]}$. 
	By using the definition of the constant $c$ as stated in this Lemma,
	we can move $|v_{t-i}|$ into the corresponding sum and thus obtain (\ref{eq:leF_wV1}) for all $t \in \mathbb{I}_{\geq0}$ and all $i \in \mathbb{I}_{[1,\mathcal{N}]}$. 
	Note that the latter step results in a lower bound of summation equal to $\tau = i$ (instead of $i+1$), which concludes this proof.
\end{IEEEproof}

\subsection{Non-discounted cost function}\label{sec:OT_ND}

We again first consider the case of the non-discounted cost function (\ref{eq:costfunction}) and provide the following result.

\begin{thm}\label{theorem:NL_3}
	Suppose that system (\ref{eq:system}) is e-IOSS and that Assumptions \ref{ass:lipschitz}, \ref{ass:cost}, \ref{ass:aux_obs}, \ref{ass:add_dist}, and \ref{ass:fullorder} apply. Choose $N\in \mathbb{I}_{\geq 1}$ arbitrarily.	
	Then, there exists $T \in \mathbb{I}_{> N}$ large enough such that the suboptimal moving horizon estimator from Definition~\ref{def:estimator} using the cost function from Definition~\ref{def:costfunction} and the candidate solution (\ref{eq:candidate}) is RGES.
\end{thm}

\begin{IEEEproof}
	\textit{Part I.} We again start with (\ref{eq:estimator}). By applying (\ref{eq:cost}), the candidate solution (\ref{eq:candidate}), prior (\ref{eq:prior}), and Assumption~\ref{ass:fullorder}, it follows that
	\begin{align} 
	J_{\text{nd}}(\hat{x}_{t-\mathcal{N}|t}, \hat{\boldsymbol{w}}_t, \hat{\boldsymbol{v}}_t) 
	&\leq
	\sum_{i = 1}^{\mathcal{N}}\big( \overline{c}_w|\tilde{w}_{t-i|t}|^a + \overline{c}_y|y_{t-i}-\tilde{y}_{t-i|t}|^a \big)  \notag \\
	&\leq (\overline{c}_w\kappa^a+\overline{c}_y)\sum_{i = 1}^{\mathcal{N}}|y_{t-i}-\tilde{y}_{t-i|t}|^a \label{eq:lem_J_0_add}
	\end{align}
	for all $t \in \mathbb{I}_{\geq 0}$.
	The rest of this proof is a straightforward modification of the proof of Lemma~\ref{lem:boundedness_of_J}, 	
	where we use Lemma~\ref{lem:boundedness_of_V} instead of Lemma~\ref{lem:boundedness_of_v} in order to upper bound the fitting error $|y_{t-i} - \tilde{y}_{t-i|t}|$. Hence it follows that (\ref{eq:lem_J}) holds for all $t \in \mathbb{I}_{\geq 0}$ with 
	\begin{equation} \label{eq:bar_sigma}
	\bar{\sigma}_{\mathcal{N}}=(\overline{c}_w\kappa^a+\overline{c}_y)(Hc)^a\frac{1-\rho^{-a\mathcal{N}}}{\rho^{a}-1}.
	\end{equation} 
	\textit{Part II.} Following the same arguments as in the proof of Lemma~\ref{lem:boundedness_of_e}, we can show that the result of Lemma~\ref{lem:boundedness_of_e} holds with the constants defined in (\ref{eq:gains_N}), where $\bar{\sigma}_{\mathcal{N}}$ is now from (\ref{eq:bar_sigma}).\\
	\textit{Part III.} By choosing $C_1$, $T$, and $\lambda$ as stated in Theorem~\ref{theorem:NL_1}, we can apply the same steps as in the proof of Theorem~\ref{theorem:NL_1}, which yields the desired result. 
\end{IEEEproof}

\subsection{Time-discounted cost function}\label{sec:OT_TD}

We now consider the case of the time-discounted cost function (\ref{eq:costfunction_td}) and provide the following result.

\begin{thm}\label{theorem:NL_4}
	Suppose that system (\ref{eq:system}) is e-IOSS and that Assumptions \ref{ass:lipschitz}, \ref{ass:cost_td}, \ref{ass:aux_obs}, \ref{ass:add_dist}, and \ref{ass:fullorder} apply. Choose $N\in \mathbb{I}_{\geq 1}$ arbitrarily.	
	Then, there exists $T \in \mathbb{I}_{> N}$ large enough such that the suboptimal moving horizon estimator from Definition~\ref{def:estimator} using the cost function from Definition~\ref{def:costfunction_td} and the candidate solution (\ref{eq:candidate}) is RGES.
\end{thm}

\begin{table*}[ht]
	\centering
	\caption{Comparison of the main characteristics of the different MHE setups considered in Sections \ref{sec:candidate_nominal}-\ref{sec:candidate_observer}.}
	\setlength\tabcolsep{5pt} 
	\begin{tabular*}{\textwidth}{@{\extracolsep{\fill}}ccccccc}
		\toprule
		\ \ Candidate solution & System dynamics & Auxiliary observer & Sec. & Cost function & Formula for $\bar{\sigma}_{\mathcal{N}}$ & Formula for $C_{\mathcal{N},1}$\\
		\midrule
		\multirow{2}{*}{\ \ Nominal traj. (\ref{eq:candidate_NL})} 
		& \multirow{2}{*}{$f(x,u,w)$}
		& \multirow{2}{*}{$\Psi_t(z_0,\boldsymbol{u},\boldsymbol{y})$}
		& \ref{sec:NT_ND}
		& n.-d. (\ref{eq:costfunction})
		& $\bar{c}_y(HC(F/\rho)^{\mathcal{N}})^a \frac{1-F^{-a\mathcal{N}}}{F^a-1}$ 
		& $(c_p(\eta/\rho)^\mathcal{N} + (c_\mathcal{N} \bar{\sigma}_{\mathcal{N}})^{\alpha})C_p$ \\
		&
		&
		& \ref{sec:NT_TD}
		& t.-d. (\ref{eq:costfunction_td})
		& $\bar{c}_yHC(F/\rho)^{\mathcal{N}} \frac{1-(\bar{\eta}/F)^{\mathcal{N}}}{(F/\bar{\eta})-1}$ 
		& $(c_p(\eta/\rho)^\mathcal{N} + \bar{\sigma}_{\mathcal{N}})C_p$\\ 
		\midrule
		\multirow{2}{*}{\ \ Observer traj. (\ref{eq:candidate})}
		& \multirow{2}{*}{$f_n(x,u) + w$}
		& \multirow{2}{*}{$f_n(z,u) + L$}
		& \ref{sec:OT_ND}
		& n.-d. (\ref{eq:costfunction})
		& $(\bar{c}_w\kappa^a+\bar{c}_y)(Hc)^a \frac{1-\rho^{-a\mathcal{N}}}{\rho^a-1}$ 
		& $(c_p(\eta/\rho)^\mathcal{N} + (c_\mathcal{N} \bar{\sigma}_{\mathcal{N}})^{\alpha})C_p$ \\
		& 
		&
		& \ref{sec:OT_TD}
		& t.-d. (\ref{eq:costfunction_td})
		& $(\bar{c}_w\kappa^a+\bar{c}_y)Hc \frac{1-(\bar{\eta}/\rho)^{\mathcal{N}}}{(\rho/\bar{\eta})-1}$
		& $(c_p(\eta/\rho)^\mathcal{N} + \bar{\sigma}_{\mathcal{N}})C_p$ \\ 
		\bottomrule
	\end{tabular*}	
	\label{tab:MHE}
\end{table*}

\begin{IEEEproof} 
	\textit{Part I.} We again start with (\ref{eq:estimator}). By applying the cost function (\ref{eq:costfunction}) with respect to Assumption~\ref{ass:cost_td}, prior (\ref{eq:prior}), candidate solution (\ref{eq:candidate}), and Assumption~\ref{ass:fullorder}, it follows that
	\begin{equation} 
	J_{\text{td}}(\hat{x}_{t-\mathcal{N}|t}, \hat{\boldsymbol{w}}_t, \hat{\boldsymbol{v}}_t) 
	\leq (\bar{c}_w\kappa+\bar{c}_y)\sum_{i = 1}^{\mathcal{N}}\bar{\eta}^{i}|y_{t-i}-\tilde{y}_{t-i|t}| \label{eq:lem_J_0_td}
	\end{equation}
	for all $t \in \mathbb{I}_{\geq 0}$.
	Applying Lemma~\ref{lem:boundedness_of_V} yields
	\begin{align*} 
	&J_{\text{td}}(\hat{x}_{t-\mathcal{N}|t}, \hat{\boldsymbol{w}}_t, \hat{\boldsymbol{v}}_t)  \leq  (\bar{c}_w\kappa+\bar{c}_y)Hc\sum_{i = 1}^{\mathcal{N}}\bar{\eta}^{i}
	\rho^{-i} \\
	& \times
	\big( C_p|x_{t-\mathcal{T}} - \hat{x}_{t-\mathcal{T}}|\rho^{\mathcal{T}} + \sum_{\tau=i}^\mathcal{T} \rho^{\tau}(C_w|w_{t-\tau}| + C_v|v_{t-\tau}|) \big).
	\end{align*}	
	Now we apply the similar steps that followed on (\ref{eq:lem_J_2}).
	In particular, we enlarge the inner sum by changing the lower bound of summation to $\tau=1$, define 
	\begin{equation} \label{eq:bar_sigma_td}
	\bar{\sigma}_{\mathcal{N}}=	(\bar{c}_w\kappa+\bar{c}_y)Hc
	\times 
	\begin{cases}
	\frac{1-(\bar{\eta}/\rho)^{\mathcal{N}}}{(\rho/\bar{\eta})-1}, & \bar{\eta} \neq \rho \\
	\mathcal{N}, & \bar{\eta} = \rho
	\end{cases}
	\end{equation} 
	and thus we have that (\ref{eq:lem_J}) holds for all $t \in \mathbb{I}_{\geq0}$ with $J_{\text{nd}}$ replaced by $J_{\text{td}}$, $a=1$, and where $\bar{\sigma}_{\mathcal{N}}$ is now from (\ref{eq:bar_sigma_td}). \\
	\textit{Part II.} Applying similar steps as in the proof of Theorem~\ref{theorem:NL_2} Part II, we can show that the result of Lemma~\ref{lem:boundedness_of_e} holds also for this case for all $t \in \mathbb{I}_{\geq0}$ with the constants defined in (\ref{eq:gains_d1}) using $\bar{\sigma}_{\mathcal{N}}$ from (\ref{eq:bar_sigma_td}). \\
	\textit{Part III.} By choosing $C_1$, $T$, and $\lambda$ as stated in Theorem~\ref{theorem:NL_1}, we can follow the same steps as in the proof of Theorem~\ref{theorem:NL_1}, which yields the desired result. 
\end{IEEEproof}

\begin{rem}\label{rem:FIE}
	We point out that, if $\bar{\eta}$ is chosen such that $\bar{\eta}<\rho$ holds\footnote{This can easily be satisfied by choosing $\bar{\eta}=\eta$ if $\rho > \eta$. If the latter condition is not satisfied (recall that we also allow for the case $\rho = \eta$ in Assumption~\ref{ass:aux_obs}), choose some $\bar{\rho} \in (\eta,1)$ and replace every $\rho$ by~$\bar{\rho}$.}, then $(\bar{\eta}/\rho)^{\mathcal{N}} \leq (\bar{\eta}/\rho)$, and hence $\bar\sigma_N$ in (\ref{eq:bar_sigma_td}) can be upper bounded by $(\bar{c}_w\kappa+\bar{c}_y)Hc(1-\bar{\eta}/\rho)/(\rho/\bar{\eta} - 1)$. Note that this upper bound on $\bar\sigma_{\mathcal{N}}$ is independent of $N$ and thus results in a bound on the estimation error that is uniformly valid for all $N$.
\end{rem}

We also note the following corollary from Theorem~\ref{theorem:NL_4}.

\begin{cor}
	Suppose that system (\ref{eq:system}) is e-IOSS and that Assumptions \ref{ass:lipschitz}, \ref{ass:cost_td}, \ref{ass:aux_obs}, \ref{ass:add_dist}, and \ref{ass:fullorder} apply. 
	If additionally $\bar{\eta}<\rho$ holds, then the (suboptimal) full information estimator (i.e., the estimator from Definition \ref{def:estimator} with $N=t$) using the time-discounted cost function (\ref{eq:costfunction_td}) and the candidate solution (\ref{eq:candidate}) with $T=N=t$ is RGES.
\end{cor}
\begin{IEEEproof}
	This follows directly from the choice of $\bar{\eta}<\rho$.
	Using the bound on $\sigma_{\mathcal{N}}$ as suggested in Remark~\ref{rem:FIE} implies that the constants in (\ref{eq:gains_d1}) are now independent of $N$.
	As a result, (\ref{eq:lem_e}) with $\mathcal{T}=\mathcal{N}=t$ for all $t \in \mathbb{I}_{\geq0}$ provides a valid bound on the estimation error of the (suboptimal) full information estimator, which finishes this proof.
\end{IEEEproof}

\section{{Discussion of theoretical properties}}\label{sec:discussion}
Table \ref{tab:MHE} summarizes the main characteristics of the MHE setups presented in Sections~\ref{sec:candidate_nominal} and \ref{sec:candidate_observer}. 
For reasons of compactness, the formulas for $\bar{\sigma}_{\mathcal{N}}$ and $C_{\mathcal{N},1}$ are given for the cases $F>1$ and $\bar{\eta}\neq\rho$ only. 
Note that we also omit the detailed description of the gains $C_1,C_2,C_3$, which are, for each case, basically similar in structure and exhibit the same qualitative behavior as $C_{\mathcal{N},1}$; a more comprehensive definition of all variables can be found in the corresponding sections.
As can be seen from the first row of Table~\ref{tab:MHE}, the use of the candidate solution~\eqref{eq:candidate_NL} allows for a very general nonlinear setup referring to the description of both system~\eqref{eq:system} and auxiliary observer~\eqref{eq:observer}.
However, since only one particular state estimate of the auxiliary observer (at time $t-\mathcal{T}$) is taken into account and the nominal dynamics are employed to construct the candidate solution, {$\bar{\sigma}_{\mathcal{N}}$ in~\eqref{eq:barsigmaN} contains the factor $F^N$; consequently, $\bar{\sigma}_{\mathcal{N}}$ and thus $C_{\mathcal{N},1}$} are exponentially increasing in $N$.
Employing the non-discounted cost function~\eqref{eq:costfunction} then introduces an additional factor $c_{\mathcal{N}}$ in the definition of $C_{\mathcal{N},1}$, resulting from Jensen's inequality applied in (\ref{eq:lem_e_5}).
This negative effect can be avoided by using the time-discounted cost function (\ref{eq:costfunction_td}) due to the more direct link between the cost function and the detectability condition in this case.
However, the exponential increase in the disturbance gains with $N$ resulting from the candidate solution remains.

By strengthening the requirements on the general setting (assuming additivity of the disturbance $w$, cf. Assumption~\ref{ass:add_dist}, and the existence of a full-order state observer, cf. Assumption~\ref{ass:fullorder}), we can construct a more sophisticated candidate solution (\ref{eq:candidate}) based on the entire trajectory of the auxiliary observer within the estimation horizon. 
Since more recent observer estimates are thus also taken into account, we can avoid the repeated use of the Lipschitz property of~$f$.
However, in the case of the non-discounted cost function (\ref{eq:costfunction}), we still obtain an exponential increase in the disturbance gains with $N$.
Note that this is due to the fact that we aim to establish exponential discounting of the disturbances in (\ref{eq:RGES}), which only became possible by applying the steps that lead to (\ref{eq:bar_sigma}), yielding gains in (\ref{eq:gains_d1}) {involving} the factor $\rho^{-\mathcal{N}}$. 
This can be overcome by using the time-discounted cost function~(\ref{eq:costfunction_td}), since here the negative effect resulting from the inverse of $\rho$ can be eliminated by a suitably chosen discount factor $\bar{\eta}$ in (\ref{eq:bar_sigma_td}), cf. Remark~\ref{rem:FIE}. 
Consequently, in this case there exist disturbance gains independent of $N$.

In the next section, we extend the results from the previous sections to allow for incorporating state and output constraints.

\section{Incorporating state and output constraints}\label{sec:constraints}

Until now, we assumed the sets $\mathbb{X}$ and $\mathbb{Y}$ to be unbounded in order to ensure feasibility of the candidate solutions in Sections~\ref{sec:candidate_nominal} and \ref{sec:candidate_observer}.
However, if the system inherently satisfies some known state and output constraints due to its physical nature (such as mechanically imposed limits on a joint angle or a measurement device, or non-negativity constraints on concentrations as in the example in Section~\ref{sec:simulation}), better results can be obtained in practice if these constraints are incorporated into the MHE problem (\ref{eq:optiMHE})-(\ref{eq:constraints}), compare \cite[Sec. 4.4]{Rawlings2017}.
In the following, we first assume that (\ref{eq:system}) and its corresponding nominal equivalent evolve in the a priori known sets $\mathbb{X} \subseteq \mathbb{R}^{n_x}$ and $\mathbb{Y} \subseteq \mathbb{R}^{n_y}$ with $\mathbb{X}$ and $\mathbb{Y}$ convex, and second, constraint satisfaction of the state implies constraint satisfaction of the output.
We show that under these conditions and by suitably adapting the proofs of the previous results, the stability guarantees remain valid (at least asymptotically) for both candidate solutions and both cost functions.

However, taking these new constraints into account requires some modifications of the candidate solutions (\ref{eq:candidate_NL}) and (\ref{eq:candidate}), as there is no guarantee that the auxiliary observer (\ref{eq:observer}) will fully satisfy them.
To this end, one could use specific observer designs such as \cite{Astolfi2021} that ensure constraint satisfaction through the use of projections.
However, this severely limits the set of possible auxiliary observers to a particular method and does not allow for user-defined customization.
To avoid this, we directly employ the projection function $\mathrm{p}_{\mathbb{X}} : \mathbb{R}^{n_x} \rightarrow \mathbb{X}$ to project the observer state $z_t$ onto the feasible set $\mathbb{X}$.
As a result, we can still consider the general class of auxiliary observers given by (\ref{eq:observer}), and thus allow for any observer design that produces robustly stable estimates irrespective of constraints.

Given the auxiliary observer at time $t \in \mathbb{I}_{\geq0}$, we denote the difference between the observer estimate at time $t-i$ for $i \in \mathbb{I}_{[1,\mathcal{N}]}$ and its projection $z_{t-i|t}-\mathrm{p}_{\mathbb{X}}(z_{t-i|t})=:\varepsilon_{t-i|t}$ as the \textit{projection error}. 
Note that $\varepsilon_{t-i|t} = 0$ if $z_{t-i|t} \in \mathbb{X}$. We aim to show the following property of the proposed suboptimal estimator.

\begin{defn}[$\boldsymbol{\varepsilon}$-RGES] \label{def:RGES_proj}
	A (suboptimal) moving horizon estimator for system~(\ref{eq:system}) is  $\varepsilon$-RGES if there exist constants $C_1,$ $C_2,$ $C_3,$ $C_{\varepsilon}>0$ and $\lambda \in (0,1)$ such that the corresponding estimate $\hat{x}_t$ at time $t \in \mathbb{I}_{\geq0}$ satisfies
	\begin{align}\label{eq:eps_RGES}
	&|x_{t}-\hat{x}_{t}| \leq C_1|x_{0}-\hat{x}_{0}|\lambda^{t} \\
	& \ \ + \sum_{\tau=1}^{t} \lambda^{\tau}
	\big( C_2^{}|w_{t-\tau}| + C_3^{}|v_{t-\tau}| \notag 
	+ C_{\varepsilon}|\varepsilon_{t-\tau|t-j}| \big)
	\end{align}
	for all initial conditions $x_0,\hat{x}_0 \in \mathbb{X}$ and all disturbance sequences $\boldsymbol{w} \in \mathbb{W}^{\infty}$ and $\boldsymbol{v} \in \mathbb{V}^{\infty}$, where $j := \lfloor \tau/T \rfloor T$.
\end{defn}

\begin{rem}
	Condition (\ref{eq:eps_RGES}) defines a slightly modified version of the stability notion given in Defition~\ref{def:RGES} that incorporates an additional disturbance term induced by the projection error~$\varepsilon$.
	If satisfied, it directly reveals that the influence of the projection error is bounded and decays over time.	
	Note that by Assumption~\ref{ass:aux_obs}, the estimation error of the observer converges to a neighborhood of the origin for $t \rightarrow \infty$. 
	Hence, if the true system state evolves in the interior of $\mathbb{X}$ and if $\mathbb{W}$ and $\mathbb{V}$ are small enough, there exists some $t^*$ such that $z_t \in \mathbb{X}$ for all $t \in \mathbb{I}_{\geq t^*}$.
	Consequently, in this case the influence of the projection error converges to zero for $t\rightarrow\infty$.
	Note also that, since we treat the difference between the observer estimate and its projection as an additional disturbance in (\ref{eq:eps_RGES}), the theoretical bound on the estimation error for suboptimal MHE gets worse when incorporating state constraints.
	In practice, however, better results can be expected \cite[Sec. 4.4]{Rawlings2017}, especially in combination with the proposed re-initialization strategy of the auxiliary observer, as can also be seen in the example in Section~\ref{sec:simulation}.
\end{rem}

We now modify both candidate solutions from Sections~\ref{sec:candidate_nominal} and \ref{sec:candidate_observer} by incorporating the projection function~$\mathrm{p}_{\mathbb{X}}$ to ensure constraint satisfaction.
For the first case, we initialize the nominal system using the projected observer estimate.
More precisely, we modify (\ref{eq:candidate_NL}) according to
\begin{equation}\label{eq:candidate_NL_proj}
	(\tilde{x}_{t-N|t}, \tilde{\boldsymbol{w}}_t, \tilde{\boldsymbol{v}}_t )
	= (\mathrm{p}_{\mathbb{X}}(z_{t-\mathcal{N}|t}), \boldsymbol{0}, \boldsymbol{0}),
\end{equation}
the prior (\ref{eq:prior}) according to
\begin{equation}\label{eq:prior_proj}
\bar{x}_{t-\mathcal{N}} = \mathrm{p}_{\mathbb{X}}(z_{t-\mathcal{N}|t}),
\end{equation}
and state the following result.

\begin{thm}\label{theorem:proj_NL}
	Suppose that system (\ref{eq:system}) is e-IOSS and that Assumptions \ref{ass:lipschitz}, \ref{ass:aux_obs}, and \ref{ass:lipschitz_f} apply. 	
	Choose some $N\in \mathbb{I}_{\geq 1}$ arbitrarily and either the non-discounted cost function (\ref{eq:costfunction}) under Assumption \ref{ass:cost} or the time-discounted cost function (\ref{eq:costfunction_td}) under Assumption \ref{ass:cost_td}.
	Let $x_0, \bar{x}_0 \in \mathbb{X}$.
	Then, there exists $T \in \mathbb{I}_{> N}$ and constants $C_1, C_2, C_3, C_{\varepsilon}>0$ and $\lambda \in (0,1)$ such that the suboptimal moving horizon estimator from Definition~\ref{def:estimator} using the candidate solution (\ref{eq:candidate_NL_proj}) is $\varepsilon$-RGES.
\end{thm}

\begin{IEEEproof}
	\textit{Part I.} We first consider the non-discounted cost function (\ref{eq:costfunction}).
	We start by following similar arguments that were needed to derive Lemma~\ref{lem:boundedness_of_v}, apply the triangle inequality to the first term of the right-hand side of (\ref{eq:leF_wv_2}) and thus obtain
	\begin{align*}
	|x_{t-\mathcal{N}}-\tilde{x}_{t-\mathcal{N}|t}| 
	&\leq |x_{t-\mathcal{N}}-z_{t-\mathcal{N}|t}| + |z_{t-\mathcal{N}|t} - \tilde{x}_{t-\mathcal{N}|t}| \\
	&= |x_{t-\mathcal{N}}-z_{t-\mathcal{N}|t}| + |\varepsilon_{t-\mathcal{N}|t}|
	\end{align*}
	for $t \in \mathbb{I}_{\geq 0}$. 
	Applying similar steps that followed (\ref{eq:leF_wv_2}), observe that (\ref{eq:leF_wv}) is modified to
	\begin{align} \label{lem:boundedness_of_v_proj}
	&|y_{t-i} - \tilde{y}_{t-i|t}| 
	\leq \sigma_{\mathcal{N}} F^{-i}
	\Big(
	C_p|x_{t-\mathcal{T}}-\hat{x}_{t-\mathcal{T}}|\rho^{\mathcal{T}} \\
	& 
	+  \sum_{\tau=i}^{\mathcal{T}}\rho^{\tau}\big(C_w|w_{t-\tau}| 
	+ C_v |v_{t-\tau}|\big) + \rho^{\mathcal{N}}C^{-1}|\varepsilon_{t-\mathcal{N}|t}| 
	\Big)  \notag
	\end{align} 
	for $t \in \mathbb{I}_{\geq 0}$  and $i \in \mathbb{I}_{[1,\mathcal{N}]}$. 
	Performing similar steps as in the proofs of Lemma~\ref{lem:boundedness_of_J} and \ref{lem:boundedness_of_e} using (\ref{lem:boundedness_of_v_proj}), and by noting that
	\begin{equation}\label{eq:prior_triangle}
	|x_{t-\mathcal{N}}-\bar{x}_{t-\mathcal{N}}| \leq |x_{t-\mathcal{N}}-z_{t-\mathcal{N}|t}| + |\varepsilon_{t-\mathcal{N}|t}|
	\end{equation} 
	in (\ref{eq:lem_e_1}) using (\ref{eq:prior_proj}) and the triangle inequality, observe that (\ref{eq:lem_e}) can then be modified to
	\begin{align} 
	&|x_{t}-\hat{x}_{t}| \leq
	C_{\mathcal{N},1}|x_{t-\mathcal{T}}-\hat{x}_{t-\mathcal{T}}|\rho^{\mathcal{T}} \notag\\
	& \ \ \ \ \ +  \sum_{\tau=1}^{\mathcal{T}} \rho^{\tau}
	\big(C_{\mathcal{N},2}|w_{t-\tau|t}|	+ C_{\mathcal{N},3}|v_{t-\tau|t}|\big) + C_{\varepsilon}'|\varepsilon_{t-\mathcal{N}|t}| \notag\\
	& \leq
	C_{\mathcal{N},1}|x_{t-\mathcal{T}}-\hat{x}_{t-\mathcal{T}}|\rho^{\mathcal{T}} \label{eq:proj_A} \\
	& \ \ \ \ \ +  \sum_{\tau=1}^{\mathcal{T}} \rho^{\tau}
	\big(C_{\mathcal{N},2}|w_{t-\tau|t}|	+ C_{\mathcal{N},3}|v_{t-\tau|t}| + C_{\varepsilon}^*|\varepsilon_{t-\tau|t}|\big) \notag
	\end{align}
	with\footnote{Note that the last step applied in (\ref{eq:proj_A}) is indeed conservative and could also be avoided to obtain a less restrictive bound on the estimation error compared to (\ref{eq:eps_RGES}). However, this step allows for a much simpler notation, since an inequality similar to (\ref{eq:proj_A}) naturally results when using the observer-based candidate solution, which is shown in the subsequent theorem.} $C_{\mathcal{N},i}$ for $i = \lbrace 1,2,3 \rbrace$ from (\ref{eq:gains_N}), $C_{\varepsilon}' := c_p\eta^\mathcal{N}+(c_\mathcal{N}\bar{\sigma}_{\mathcal{N}})^{\alpha}\rho^\mathcal{N}C^{-1}$, $C_{\varepsilon}^* := C_{\varepsilon}'\rho^{-\mathcal{N}}$, and with $\bar{\sigma}_{\mathcal{N}}$ according to (\ref{eq:barsigmaN}). 
	Now, by choosing $T, \lambda$, and $C_i$ for $i = \lbrace 1,2,3 \rbrace$ as in Theorem~\ref{theorem:NL_1} and its proof, we obtain
	\begin{align*} 
	|x_{t}-\hat{x}_{t}| & \leq
	|x_{t-\mathcal{T}}-\hat{x}_{t-\mathcal{T}}|\lambda^{\mathcal{T}} \\
	& \ \ \ +  \sum_{\tau=1}^{\mathcal{T}} \lambda^{\tau}
	\big(C_2|w_{t-\tau|t}|	+ C_3|v_{t-\tau|t}| + C_{\varepsilon}|\varepsilon_{t-\tau|t}|\big),
	\end{align*}
	where $C_{\varepsilon}:= C_1^{-1/T}\max_{\mathcal{N} \in \mathbb{I}_{[0,N]}}C_{\varepsilon}^*$.
	Performing similar steps as in the proof of Theorem~\ref{theorem:NL_1} then yields (\ref{eq:eps_RGES}).\\
	\textit{Part II.} We now consider the time-discounted cost function~\eqref{eq:costfunction_td}. 
	We start with (\ref{eq:lem_J_0}) and the modified version of Lemma~\ref{lem:boundedness_of_v} given in (\ref{lem:boundedness_of_v_proj}).
	By applying similar steps that followed on (\ref{eq:lem_J_0}) together with arguments from the first part of this proof, we obtain (\ref{eq:proj_A})
	with $C_{\mathcal{N},i},i = \lbrace 1,2,3 \rbrace$ from~\eqref{eq:gains_d1} and $C_{\varepsilon}^* = c_p(\eta/\rho)^\mathcal{N} + \bar{\sigma}_{\mathcal{N}}C^{-1}$.
	By applying the same steps that followed on (\ref{eq:proj_A}), we obtain (\ref{eq:eps_RGES}), which finishes this proof.
\end{IEEEproof}

We now consider the second candidate solution introduced in Section~\ref{sec:candidate_observer}.
If Assumptions~\ref{ass:add_dist} and \ref{ass:fullorder} hold, we can project the full state trajectory of the observer within the estimation horizon onto the feasible set $\mathbb{X}$, which yields $\tilde{x}_{t-i|t} = \mathrm{p}_{\mathbb{X}}(z_{t-i|t})$ for all $t \in \mathbb{I}_{\geq 0}$ and $i \in \mathbb{I}_{[1,\mathcal{N}]}$.
To obtain some $\tilde{w}_{t-i|t}$ such that (\ref{eq:system_a}) (with respect to Assumption~\ref{ass:add_dist}) is satisfied, we first note that $\tilde{x}^+ = f_n(\tilde{x},u) + \tilde{w} = \mathrm{p}_{\mathbb{X}}(z^+)$ and hence $\tilde{w} = \mathrm{p}_{\mathbb{X}}(z^+) - f_n(\tilde{x},u)$, again exploiting one-step controllability with respect to the input $\tilde{w}$.
We therefore modify (\ref{eq:candidate}) to 
\begin{subequations}
	\label{eq:candidate_proj}
	\begin{align}
	&(\tilde{x}_{t-\mathcal{N}|t}, \tilde{\boldsymbol{w}}_t, \tilde{\boldsymbol{v}}_t ) = 
	(\mathrm{p}_{\mathbb{X}}(z_{t-\mathcal{N}|t}), \tilde{\boldsymbol{w}}_t, \boldsymbol{0}),\\
	& \tilde{w}_{t-i|t} =  \mathrm{p}_{\mathbb{X}}(z_{t-i+1|t}) - f_n(\tilde{x}_{t-i|t},u_{t-i}), \ \ i \in \mathbb{I}_{[1,\mathcal{N}]} 	\label{eq:candidate_proj_b}
	\end{align}
\end{subequations}
and provide the corresponding result.

\begin{thm}\label{theorem:proj}
	Suppose that system (\ref{eq:system}) is e-IOSS and that Assumptions \ref{ass:lipschitz}, \ref{ass:aux_obs}-\ref{ass:fullorder} apply. 	
	Choose some $N\in \mathbb{I}_{\geq 1}$ arbitrarily and either the non-discounted cost function~(\ref{eq:costfunction}) under Assumption \ref{ass:cost} or the time-discounted cost function (\ref{eq:costfunction_td}) under Assumption \ref{ass:cost_td}.
	Let $x_0, \bar{x}_0 \in \mathbb{X}$.
	Then, there exists $T \in \mathbb{I}_{> N}$ and constants $C_1, C_2, C_3, C_{\varepsilon}>0$ and $\lambda \in (0,1)$ such that the suboptimal moving horizon estimator from Definition~\ref{def:estimator} using the candidate solution (\ref{eq:candidate_proj}) is \mbox{$\varepsilon$-RGES}.
\end{thm}

\begin{IEEEproof}
	\textit{Part I.} 
	Using (\ref{eq:candidate_proj}), (\ref{eq:system_b}), the triangle inequality and Assumption~\ref{ass:lipschitz}, the fitting error of the candidate solution can be bounded by
	\begin{equation}
	|y_{t-i}-\tilde{y}_{t-i|t}| \leq  |y_{t-i}-h_n(z_{t-i|t},u_{t-i})| + H|\varepsilon_{t-i|t}|. \label{eq:proj_a}
	\end{equation}
	To construct a similar bound on $\tilde{w}_{t-i|t}$, we first note that for a given $a \in \mathbb{R}^{n_x}$, $|\mathrm{p}_{\mathbb{X}}(a)-b| \leq |a-b|$ for any $b \in \mathbb{X}$, since by convexity of $\mathbb{X}$ and optimality of $\mathrm{p}_{\mathbb{X}}$, the angle between $\mathrm{p}_{\mathbb{X}}(a)-a$ and $a-b$ is obtuse \cite[Thm. 3.1.1]{HiriartUrruty1993}.
	Now consider (\ref{eq:candidate_proj_b}) and recall that $f_n(\tilde{x}_{t-i|t},u_{t-i})\in\mathbb{X}$.
	Application of Assumptions~\ref{ass:lipschitz_f} and \ref{ass:fullorder} then yields
	\begin{align}
		|\tilde{w}_{t-i|t}| &= |\mathrm{p}_{\mathbb{X}}(z_{t-i+1|t}) -  f_n(\tilde{x}_{t-i|t},u_{t-i})| \notag\\
		& \leq |z_{t-i+1|t} - f_n(\tilde{x}_{t-i|t},u_{t-i})| \notag\\
		& = |f_n(z_{t-i|t},u_{t-i}) + L_{t-i|t} - f_n(\tilde{x}_{t-i|t},u_{t-i})|\notag\\
		& \leq F|z_{t-i|t}-\tilde{x}_{t-i|t}| + |L_{t-i|t}| \notag\\
		& \leq F|\varepsilon_{t-i|t}| + \kappa|y_{t-i}-h_n(z_{t-i|t},u_{t-i})|  \label{eq:proj_b}
	\end{align}
	for $t \in \mathbb{I}_{\geq 0}$ and $i \in \mathbb{I}_{[1,\mathcal{N}]}$.
	Using (\ref{eq:proj_a}), (\ref{eq:proj_b}), and by applying Jensen's inequality, it follows that (\ref{eq:lem_J_0_add}) is modified to
	\begin{align} 
		J_{\text{nd}}(&\hat{x}_{t-\mathcal{N}|t}, \hat{\boldsymbol{w}}_t, \hat{\boldsymbol{v}}_t) \label{eq:proj_b1} \\
		&\leq \sigma_1\sum_{i = 1}^{\mathcal{N}}  |y_{t-i}-h_n(z_{t-i|t},u_{t-i})|^a +
		\sigma_2\sum_{i = 1}^{\mathcal{N}} |\varepsilon_{t-i|t}|^a , \notag
	\end{align}
	where $\sigma_1: = \overline{c}_w\kappa(F+\kappa)^{a-1}+\overline{c}_y(H+1)^{a-1}$ and $\sigma_2: = \overline{c}_wF(F+\kappa)^{a-1}+\overline{c}_yH(H+1)^{a-1}$.	
	Now we exploit that Lemma~\ref{lem:boundedness_of_V} provides a valid bound on the output differences in (\ref{eq:proj_b1}).
	By performing the similar steps that followed on (\ref{eq:lem_J_0_add}) and using the fact that $\sum_ir_i^a \leq (\sum_ir_i)^a$ for $r_i\geq0$ and $a\geq1$, we obtain
	\begin{align} 
	&J_{\text{nd}}(\hat{x}_{t-\mathcal{N}|t}, \hat{\boldsymbol{w}}_t, \hat{\boldsymbol{v}}_t) 
	\leq \sigma_1\bar{\sigma}_{\mathcal{N}} \Big(C_p|x_{t-\mathcal{T}} - \hat{x}_{t-\mathcal{T}}|\rho^{\mathcal{T}} \label{eq:proj_c} \\
	& \ \ \ + \sum_{\tau=i}^\mathcal{T} \rho^{\tau}\big(C_w|w_{t-\tau}| + C_v|v_{t-\tau}|\big) \Big)^a + \sigma_2\Big(\sum_{i = 1}^{\mathcal{N}}|\varepsilon_{t-i|t}| \Big)^a, \notag
	\end{align}	
	where $\bar{\sigma}_{\mathcal{N}} = (Hc)^a(1-\rho^{-a\mathcal{N}})/(\rho^{a}-1)$, compare (\ref{eq:bar_sigma}).
	We proceed similar as in the proof of Lemma~\ref{lem:boundedness_of_e} and consider (\ref{eq:lem_e_5}) together with (\ref{eq:proj_c}).
	Using that\footnote{This is true since $r\rightarrow r^{\alpha}$ strictly increases for all $x\geq0$ and therefore is a $\mathcal{K}$-function. For a general proof, see \cite{Rawlings2012}.} $(a+b)^{\alpha} \leq (2a)^{\alpha} + (2b)^{\alpha}$ for $a,b \geq0$ leads to
	\begin{align*}
	&\underline{c}_p\eta^\mathcal{N}|\hat{x}_{t-\mathcal{N}|t} - \bar{x}_{t-\mathcal{N}}|
	+ \sum_{i=1}^\mathcal{N}\eta^i\big(\underline{c}_w|\hat{w}_{t-i|t}| + \underline{c}_v|\hat{v}_{t-i|t}| \\ 
	& + \underline{c}_y|y_{t-i}-\hat{y}_{t-i|t}|\big) 
	\leq 
	(2c_\mathcal{N}\sigma_1\bar{\sigma}_{\mathcal{N}})^{\alpha} 
	\big(C_p|x_{t-\mathcal{T}} - \hat{x}_{t-\mathcal{T}}|\rho^{\mathcal{T}} \\
	& + \sum_{\tau=i}^\mathcal{T} \rho^{\tau}\big(C_w|w_{t-\tau}| + C_v|v_{t-\tau}|\big) \big) + (2c_\mathcal{N}\sigma_2)^{\alpha}\sum_{i = 1}^{\mathcal{N}}|\varepsilon_{t-i|t}|.
	\end{align*}
	By \eqref{eq:lem_e_1} together with the prior~\eqref{eq:prior_proj}, the triangle inequality~\eqref{eq:prior_triangle}, and {Lemma}~\ref{cor:obs}, we obtain~\eqref{eq:proj_A}, where
	\begin{subequations}\label{eq:gains_N_proj}
		\begin{align} 
		C_{\mathcal{N},1} &:=
		(c_p(\eta/\rho)^\mathcal{N} 
		{\medmuskip=0mu \, + \, } (2c_\mathcal{N} \sigma_1\bar{\sigma}_{\mathcal{N}})^{\alpha})C_p,\\
		C_{\mathcal{N},2} &:=
		(c_p(\eta/\rho)^\mathcal{N} 	
		{\medmuskip=0mu \, + \, } (2c_\mathcal{N} \sigma_1\bar{\sigma}_{\mathcal{N}})^{\alpha})C_w
		{\medmuskip=0mu \, + \, } c_w\eta/\rho,\\
		C_{\mathcal{N},3} &:=
		(c_p(\eta/\rho)^\mathcal{N} 
		{\medmuskip=0mu \, + \, } (2c_\mathcal{N} \sigma_1\bar{\sigma}_{\mathcal{N}})^{\alpha})C_v {\medmuskip=0mu \, + \, } c_v\eta/\rho,\\
		C_{\varepsilon}^* &:=
		c_p(\eta/\rho)^\mathcal{N} {\medmuskip=0mu \, + \, } (2c_\mathcal{N}\sigma_{2})^{\alpha}\rho^\mathcal{-N}.
		\end{align}
	\end{subequations}
	Applying the same steps that followed on (\ref{eq:proj_A}) yields (\ref{eq:eps_RGES}).\\
	\textit{Part II.} We now consider the time-discounted cost function (\ref{eq:costfunction_td}).
	By starting with the same steps as in the proof of Theorem~\ref{theorem:NL_4} using the bounds established in (\ref{eq:proj_a}) and (\ref{eq:proj_b}), observe that~\eqref{eq:lem_J_0_td} is modified to
	\begin{align*} 
	J_{\text{td}}(\hat{x}_{t-\mathcal{N}|t}, \hat{\boldsymbol{w}}_t,  \hat{\boldsymbol{v}}_t) &\leq \sum_{i = 1}^{\mathcal{N}}\bar{\eta}^{i}\big((\bar{c}_wF+\bar{c}_yH)|\varepsilon_{t-i|t}| \\ 
	& \ \ \ + 
	(\bar{c}_w\kappa+\bar{c}_y)|y_{t-i}-h_n(z_{t-i|t},u_{t-i})| \big)
	\end{align*}
	Applying Lemma~\ref{lem:boundedness_of_V} and the similar steps that followed on (\ref{eq:lem_J_0_td}) together with (\ref{eq:prior_triangle}), we can show that (\ref{eq:proj_A}) holds with $C_{\mathcal{N},i}$ for $i = \lbrace 1,2,3 \rbrace$ from Theorem~\ref{theorem:NL_4} and	where 
	$C_{\varepsilon}^*=(c_p\eta^\mathcal{N} + \bar{c}_wF+\bar{c}_yH)\max_{i \in \lbrace 1,\mathcal{N}\rbrace}(\bar{\eta}/\rho)^i$.
	Applying the same steps that followed on (\ref{eq:proj_A}) yields (\ref{eq:eps_RGES}).
\end{IEEEproof}

Table~\ref{tab:constraints} compares the different candidate solutions from the previous sections in terms of the respective possible constraints on the domain of the estimated trajectory that are guaranteed to be satisfied.
\begin{table}[t]
	\centering
	\caption{Permissible constraints on the sets involved.}
	\begin{tabular}{cccccc}
		\toprule
		Candidate solution & Eq. &$\mathbb{X}$ & $\mathbb{W}$ & $\mathbb{V}$ & $\mathbb{Y}$ \\
		\midrule
		\multirow{2}{*}{Nominal trajectory}
		& (\ref{eq:candidate_NL})
		& $= \mathbb{R}^{n_x}$
		& $\subseteq \mathbb{R}^{n_w}$
		& $\subseteq \mathbb{R}^{n_v}$
		& $= \mathbb{R}^{n_y}$ \\
		& (\ref{eq:candidate_NL_proj})
		& $\subseteq \mathbb{R}^{n_x}$
		& $\subseteq \mathbb{R}^{n_w}$
		& $\subseteq \mathbb{R}^{n_v}$
		& $\subseteq \mathbb{R}^{n_y}$ \\	
		\midrule
		\multirow{2}{*}{Observer trajectory}
		& (\ref{eq:candidate})
		& $= \mathbb{R}^{n_x}$
		& $= \mathbb{R}^{n_w}$
		& $\subseteq \mathbb{R}^{n_v}$
		& $= \mathbb{R}^{n_y}$ \\	
		& (\ref{eq:candidate_proj})
		& $\subseteq \mathbb{R}^{n_x}$
		& $= \mathbb{R}^{n_w}$
		& $\subseteq \mathbb{R}^{n_v}$
		& $\subseteq \mathbb{R}^{n_y}$ \\
		\bottomrule
	\end{tabular}	
	\label{tab:constraints}
\end{table}
The use of the candidate solution (\ref{eq:candidate_NL_proj}) based on the projected nominal trajectory allows incorporating the most information into the optimization problem compared to the other setups by constraining all the sets $\mathbb{X}$, $\mathbb{W}$, $\mathbb{V}$, and $\mathbb{Y}$, provided that they are known a priori.
On the other hand, we have the candidate solutions based on the observer trajectory, where we need $\mathbb{W}$ radially unbounded to ensure one-step controllability in order to be able to exactly reproduce the trajectory of the auxiliary observer (or to move the observer state into the feasible set).
{Thus, we can easily incorporate state and output constraints into the MHE schemes from Sections~\ref{sec:candidate_nominal}-\ref{sec:candidate_observer}, while preserving the theoretical guarantees.}
{To conclude this section, we note that the required changes in Algorithm~\ref{alg:sMHE} (including its modifications below \eqref{eq:candidate}) to produce the current state estimate $\hat{x}_t$ are minor if constraints are to be considered.
Indeed, the respective steps remain unchanged, only the prior has to be replaced by \eqref{eq:prior_proj} and the candidate solution by either \eqref{eq:candidate_NL_proj} or \eqref{eq:candidate_proj}}.

\section{Simulation case study}\label{sec:simulation}
In order to illustrate our results, we apply the proposed estimator to the set of batch chemical reactions $A \rightleftharpoons B + C,\ 2B \rightleftharpoons C$; this example is adopted from \cite[Example 4.39]{Rawlings2017}.
The Euler-discretized model describing the evolution of the concentrations over time corresponds to
\begin{align*}
x_1^+ &= x_1 + t_\Delta(-p_1x_1+p_2x_2x_3) + w_1,\\
x_2^+ &=	x_2 + t_\Delta(p_1x_1-p_2x_2x_3-2p_3x_2^2+2p_4x_3) + w_2, \\
x_3^+ &=	x_3 + t_\Delta(p_1x_1-p_2x_2x_3+p_3x_2^2-p_4x_3) + w_3, \\
y &= x_1 + x_2 + x_3 + v,
\end{align*}
with the step size $t_\Delta = 0.25$.
We choose the parameter vector $p = [0.2, 0.05, 0.2, 0.1]$ and the initial conditions $x_0 = [0.5, 0.05, 0]^\top$ and $\bar{x}_0 = [1, 0.5, 0.1]^\top$.
We consider the prior knowledge $\mathbb{X} = \lbrace x \in \mathbb{R}^{3}_{\geq0} : x_i \leq 4 , i = \lbrace 1, 2, 3 \rbrace \rbrace$, where non-negativity follows from physical nature and the upper bound provides a realistic compact set with respect to the initial conditions.
During the simulations, the disturbances $w$ and $v$ are treated as uniformly distributed random variables which are sampled from the sets $\mathbb{W} = \lbrace w \in \mathbb{R}^{3} : |w_i| \leq 2\cdot10^{-3}, i = \lbrace 1, 2, 3 \rbrace \rbrace$ and $\mathbb{V} = \lbrace v \in \mathbb{R} : |v| \leq 10^{-2} \rbrace$.

\subsubsection*{{Auxiliary observer}} We choose a nonlinear Luenberger approach based on the multi-dimensional mean-value theorem, cf. \cite{Zemouche2006a,Ibrir2009}.
Following \cite{Zemouche2006a}, we can write $f_n(x)-f_n(z) = A(\Theta)(x-z)$ for $x,z$ evolving in some set $\overline{\mathbb{X}}$, where $A$ contains the partial derivatives of $f_n$ with respect to $x$ and the matrix $\Theta \in \mathbb{R}^{3 \times 3}$ is a time-varying parameter evolving in the convex set $\mathbb{H}$ depending on $\overline{\mathbb{X}}$.
In the following, we choose $\overline{\mathbb{X}}$ as a proper superset of $\mathbb{X}$, i.e., $\mathbb{X}  \subset \overline{\mathbb{X}} := \lbrace x \in \mathbb{R}^{3} : -0.03 \leq x_2 \leq 4, -2 \leq x_3 \leq 4 \rbrace$, since there are no guarantees that the Luenberger observer provides non-negative estimates.
Now closing the loop through the output injection law $K(h_n(z)-y)$ with $h_n(z) = Cz$ and $C = [1, 1, 1]$ (cf. Remark~\ref{rem:fullorder}), we arrive at the linear parameter-varying error equation $x^+-z^+ = (A(\Theta)+KC)(x-z) + w + Kv$.
With $P$ being a symmetric positive definite matrix, we define the \mbox{P-norm}\footnote{
	{Note that the induced norm of $|\cdot|_P$ for a matrix $A$ is given by $|A|_P = |P^{1/2}AP^{-1/2}|$.}
} $|\cdot|_P = |P^{\frac{1}{2}}\cdot|$, apply it to both sides of the error equation, and by using the triangle inequality and submultiplicativity, we obtain that
\begin{equation}
|x^+-z^+|_P \ {\leq} \ |A(\Theta)+KC|_P|x-z|_P + |w|_P + |Kv|_P. \label{eq:ex_error}
\end{equation}
To achieve exponential contraction of the observer error, we require $|A(\Theta)+KC|_P\leq\rho \in (0,1)$ for all $\Theta \in \mathbb{H}$.
{By convexity of $\mathbb{H}$ and linearity of $A(\Theta)$}, this can be guaranteed by solving the corresponding LMI\footnote{
	{LMIs were solved in Matlab using the toolbox YALMIP \cite{Loefberg2004} and the semidefinit programming solver MOSEK \cite{MOSEKApS2019}.}
} for all the vertices of $\mathbb{H}$ (cf.~\cite[Sections 1.2 and 4.3]{GuangRenDuan2013}), which is satisfied for, e.g., the decay rate $\rho = 0.985$, 
\begin{equation*}
	P = 
	{\begin{pmatrix} 
	7.231 & 3.063  & 1.957 \\
	3.063 & 35.606 & 1.746 \\
	1.957 &	1.746  & 2.705
	\end{pmatrix}}, \quad \text{and} \quad
	K = {\begin{pmatrix} -0.129 \\ -0.069 \\ -0.923 \end{pmatrix}}.
\end{equation*}
Repeated application of (\ref{eq:ex_error}) then lets us conclude that the Luenberger observer is RGES on the set $\overline{\mathbb{X}}$ and satisfies (\ref{eq:RGES_observer}) with \mbox{$C_p =4.282$}, $C_w = 4.347$, $C_v = 1.322$.

\subsubsection*{{Verification of e-IOSS}}
{Based on the robustly stable observer given above, we can now easily verify e-IOSS on the set $\overline{\mathbb{X}}$ by suitably adapting the arguments from \cite[Sec. VI]{Knuefer2020} for linear systems.
For arbitrary $x,\chi\in\overline{\mathbb{X}}$, $w,\omega\in{\mathbb{W}}$, and $v,\nu\in{\mathbb{V}}$, and with $y=h(x,v)$ and $\zeta=h(\chi,\nu)$, we have that
$x^+-\chi^+ = f_n(x)-f_n(\chi) + w - \omega$. We add $0 = K(y-\zeta) - K(y - \zeta) = KC(x - \chi) + K(v - \nu) - K(y - \zeta)$ to the right-hand side, use $f_n(x)-f_n(\chi)=A(\Theta)(x-\chi)$, apply the P-norm to both sides, exploit the observer decay rate $\rho$, and obtain
\begin{equation*}
	|x^+{-}\chi^+|_P{\,\leq\,} \rho|x{-}\chi|_P{\,+}|w{-}\omega|_P{\,+}|K(v{-}\nu)|_P{\,+}|K(y{-}\zeta)|_P
\end{equation*}
which is a Lyapunov characterization of e-IOSS, compare~\cite{Knuefer2020}.
Consequently, we can derive \eqref{eq:eIOSS} with $c_p=C_p$, $c_u=0$, $c_w = C_w$, $c_v = c_y = C_v$, and $\eta = \rho$, thus proving e-IOSS of the original system.}

\subsubsection*{{Suboptimal MHE scheme}} We design four different moving horizon estimators based on the methods presented in this paper and compare their properties and estimation results.
Since we aim to incorporate the prior knowledge about the set $\mathbb{X}$ (and $\mathbb{V}$, which can in general be considered by all the candidate solutions from Table~\ref{tab:constraints}) into the MHE problem~(\ref{eq:optiMHE})-(\ref{eq:constraints}), we focus on the designs proposed in Section~\ref{sec:constraints}.
{We choose the horizon length $N=3$ and the functions $\Gamma(\chi,\bar{x}) = c_p|\chi-\bar{x}|^a$ and $l(\omega,\nu,y-\zeta) = c_w|\omega|^a + c_v|\nu|^a + c_y|y-\zeta|^a$ with $a=2$ for the non-discounted cost function~\eqref{eq:costfunction} (which will therefore be termed as \textit{quadratic} (q.-c.) in the following) and with $a=1$ for the time-discounted (t.-d.) cost function~\eqref{eq:costfunction_td}.}
For the latter, we also choose $\bar{\eta}=\eta$, and as a result both compatibility conditions from Assumptions~\ref{ass:cost} and~\ref{ass:cost_td} hold with equality{, compare Remark~\ref{rem:cost}}.
\begin{table}
	\centering
	\caption{Disturbance gains and value of $T_{\text{min}}$ for different MHE setups.}
	\setlength\tabcolsep{5pt} 
	\begin{tabular*}{\columnwidth}{@{\extracolsep{\fill}}ccccccc}
		\toprule
		\ \ Candidate solution & Cost & $C_1$ & $C_2$ & $C_3$ & $C_{\varepsilon}$ & $T_{\text{min}}$ \\\midrule
		\multirow{2}{*}{\ \ Nom. traj. (\ref{eq:candidate_NL_proj}) } 
		& q.-c. 
		& 64.95 & 69.24 & 21.05 & 15.17 & 277 \\ 
		& t.-d. 
		& 36.55 & 40.83 & 12.42 & 8.54 & 239 \\\midrule
		\multirow{2}{*}{\ \ Obs. traj. (\ref{eq:candidate_proj})}  
		& q.-c. 
		& 193.35 & 197.64 & 60.10 & 29.37 & 349 \\
		& t.-d. 
		& 87.48 & 91.76 & 27.91 & 10.00 & 296 \\ \bottomrule
	\end{tabular*}	
	\label{tab:gains}
\end{table} 
Table~\ref{tab:gains} compares the resulting values of the disturbance gains $C_1,C_2,C_3$ and $C_{\varepsilon}$ in (\ref{eq:eps_RGES}) by following Theorems~\ref{theorem:proj_NL} and \ref{theorem:proj}, respectively.
{In line with the main observations in Section~\ref{sec:discussion}, it can be seen that the time-discounted cost function generally leads to smaller disturbance gains than the quadratic cost function.
Interestingly, we can observe that the candidate solution~\eqref{eq:candidate_NL_proj} leads to smaller gains\footnote{
	{This results from the fact that, for this specific example, both the Lipschitz constant of $f$ and the decay rate of the auxiliary observer are close to one ($F \approx 1.06$, $\rho = 0.985$), and $N$ is chosen small.
	Consequently, the influence of the term $(F/\rho)^N$ from Lemma~\ref{lem:boundedness_of_v} on Theorem~\ref{theorem:proj_NL} is smaller than the combined influence resulting from the constants $\overline{c}_w$ and $\overline{c}_y$ contained in $\bar{\sigma}_{\mathcal{N}}$ and $\sigma_1$ that are used in Theorem~\ref{theorem:proj}}.
} than candidate solution~\eqref{eq:candidate_proj}.}
{Besides, in this example we also find that the disturbance gains computed for suboptimal MHE are generally much worse than those of the auxiliary observer given above.
This is on the one hand due to the fact that we guarantee stability without any optimization (and hence the disturbance gains cannot be better than that of the auxiliary observer), and on the other hand due to various conservative steps within the respective proofs.}
{Combined with the fact that the decay rate $\rho$ is already rather conservative, the minimum values of $T$ (denoted as $T_{\min}$) required in each case to ensure robust stability according to condition~\eqref{eq:def_C1} become large, as it can be seen in the last column of Table~\ref{tab:gains}; yet good simulation results are also obtained with a much smaller $T$.
Hence, the above guarantees should rather be interpreted to be of conceptual nature.}
To illustrate the potential of the proposed re-initialization strategy in practice (cf. Remark~\ref{rem:choice_T}), we choose $T=5$ in the following, although we must note that this choice is not theoretically covered.

\subsubsection*{{Simulation results}}
\begin{figure}
	\centering
	\includegraphics{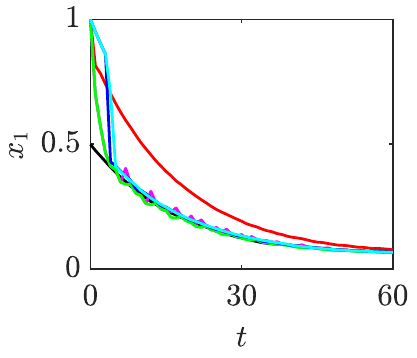}
	\includegraphics{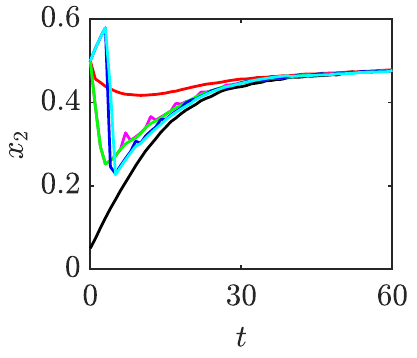} \\
	\includegraphics{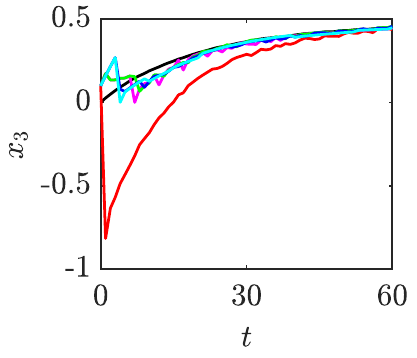}
	\includegraphics{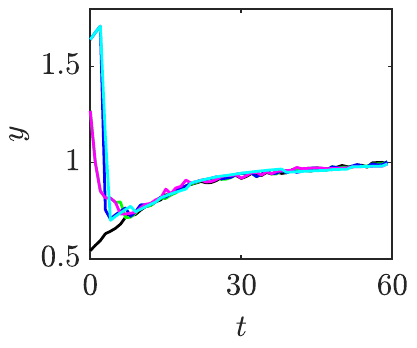}\\
	\caption{Comparison between suboptimal MHE using the candidate solution (\ref{eq:candidate_NL_proj}) with the non-discounted cost function (green) and the time-discounted cost function (blue), using the candidate solution (\ref{eq:candidate_proj}) with the non-discounted cost function (magenta) and the time-discounted cost function (cyan), the Luenberger observer (red), and the real system (black). The optimizer solving the NLP with the non-discounted (time-discounted) cost function is terminated after $i=2$ ($i = 10$) iterations.}
	\label{fig:example_xy}
\end{figure}

Figure~\ref{fig:example_xy} provides the estimation results\footnote{
	{The MHE simulations were performed on a standard PC (Intel Core i7 with 2.6 GHz, 12 MB cache, and 16 GB RAM under Ubuntu Linux 20.04) in Matlab with CasADi \cite{Andersson2018} and the NLP solver IPOPT \cite{Waechter2005}. We have employed a multiple shooting approach (cf.~\cite{Kuehl2011}), applied a re-formulation in line with Remark~\ref{rem:implementing}, and used the improved warm start from Remark~\ref{rem:decrease_condition}.}
} for the different configurations of the suboptimal estimator {compared to the real system states and the Luenberger observer that is used to construct the candidate solutions.}
This illustrates that all suboptimal estimators are robustly stable and moreover capable of improving the estimates of the auxiliary observer (that evolves outside of $\mathbb{X}$ in its transient phase) with very few iterations while ensuring constraint satisfaction.
\begin{figure}
	\centering
	\includegraphics{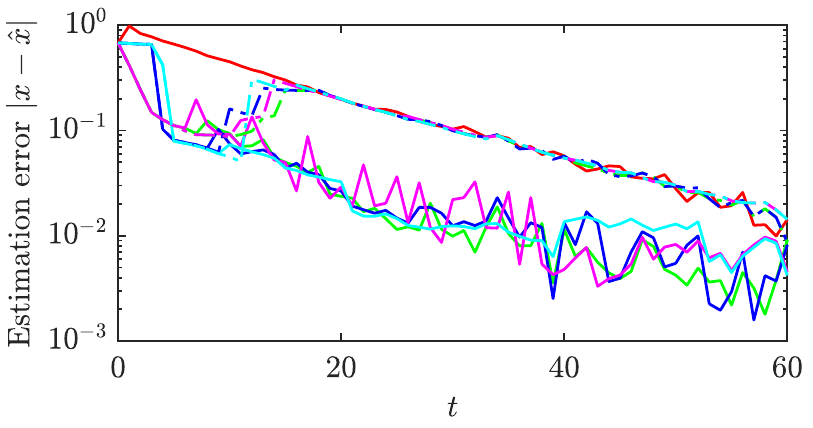} 
	\caption{Corresponding estimation error of the trajectories from Figure~\ref{fig:example_xy} (solid lines) compared to their counterparts with $T=t$, i.e., without re-initializing the auxiliary observer (dotted-dashed lines).}
	\label{fig:example_e}
\end{figure}
\begin{table*}
	\centering
	\caption{SSE, SNE, and $\tau_{\mathrm{a}}$ for the Luenberger observer compared to different settings of the proposed suboptimal estimator for different maximum numbers of allowed iterations $i$ of the optimizer. Each value represents the average over 100 simulations.}
	\setlength\tabcolsep{5pt} 
	\begin{tabular*}{\textwidth}{@{\extracolsep{\fill}} c c c cccc c cccc c cccc}
		\toprule
		\multicolumn{2}{c}{\multirow{3}{*}{Configuration} }
		& \phantom{0} & \multicolumn{4}{c}{SSE} 
		& \phantom{0} & \multicolumn{4}{c}{SNE} 
		& \phantom{0} & \multicolumn{4}{c}{$\tau_{\mathrm{a}} \ [\text{ms}]$ } 
		\\
		\cmidrule(lr){4-7}
		\cmidrule(lr){9-12}
		\cmidrule(lr){14-17}
		& &
		& \multicolumn{2}{c}{Nom. traj. (\ref{eq:candidate_NL_proj})} 
		& \multicolumn{2}{c}{Obs. traj. (\ref{eq:candidate_proj})}  
		& & \multicolumn{2}{c}{Nom. traj. (\ref{eq:candidate_NL_proj})} 
		& \multicolumn{2}{c}{Obs. traj. (\ref{eq:candidate_proj})}		
		& & \multicolumn{2}{c}{Nom. traj. (\ref{eq:candidate_NL_proj})} 
		& \multicolumn{2}{c}{Obs. traj. (\ref{eq:candidate_proj})}
		\\
		\cmidrule(lr){4-5}
		\cmidrule(lr){6-7}
		\cmidrule(lr){9-10}
		\cmidrule(lr){11-12}
		\cmidrule(lr){14-15}
		\cmidrule(lr){16-17}
		& 
		& & \ q.-c. & t.-d. & \ q.-c. & t.-d. 
		& & \ q.-c. & t.-d. & \ q.-c. & t.-d. 
		& & \ q.-c. & t.-d. & \ q.-c. & t.-d. 
		\\\midrule
		\multicolumn{2}{c}{Luenberger observer}
		& & \multicolumn{4}{c}{6.25} 
		& & \multicolumn{4}{c}{13.97} 
		& & \multicolumn{4}{c}{-} 
		\\\midrule
		\multirow{8}{*}{\shortstack{\hspace*{1ex} $N=3$\\[0.5ex] \hspace*{1.35ex} $T=5$}}
		& $i=   0$ 
		& & 3.50 & 3.50 & 2.59 & 2.60 
		& & 9.47 & 9.47 & 8.57 & 8.58 
		& &   2.06 &   2.22 &   2.05 &   2.22 
		\\ 
		& $i=   1$ 
		& & 1.97 & 3.50 & 1.94 & 3.28 
		& & 6.46 & 9.48 & 6.55 & 9.22 
		& &   2.82 &   3.04 &   2.81 &   3.04 
		\\ 
		& $i=   2$ 
		& & 0.86 & 2.15 & 0.88 & 1.73 
		& & 3.23 & 6.68 & 3.35 & 5.48 
		& &   3.55 &   3.80 &   3.55 &   3.82 
		\\ 
		& $i=   5$ 
		& & 0.84 & 3.49 & 0.84 & 3.16 
		& & 3.46 & 9.46 & 3.45 & 8.75 
		& &   5.71 &   6.34 &   5.82 &   6.41 
		\\ 
		& $i=  10$ 
		& & 0.88 & 1.90 & 0.88 & 2.03 
		& & 3.69 & 4.33 & 3.69 & 4.39 
		& &   8.09 &  10.65 &   8.23 &  10.66 
		\\ 
		& $i=  15$ 
		& & 0.88 & 1.44 & 0.88 & 1.44 
		& & 3.68 & 3.31 & 3.68 & 3.26 
		& &   8.18 &  14.55 &   8.31 &  14.54 
		\\ 
		& $i=  20$ 
		& & 0.88 & 1.36 & 0.88 & 1.36 
		& & 3.70 & 3.03 & 3.70 & 2.99 
		& &   8.18 &  18.38 &   8.31 &  18.39 
		\\ 
		& $i=  50$ 
		& & 0.88 & 1.17 & 0.88 & 1.17 
		& & 3.69 & 2.82 & 3.69 & 2.82 
		& &   8.14 &  44.16 &   8.27 &  44.17 
		\\ \midrule
		\multirow{6}{*}{\shortstack{\hspace*{1ex} $N=10$\\[0.5ex] \hspace*{1.35ex} $T=15$}}
		& $i=   1$ 
		& & 2.59 & 5.82 & 2.64 & 3.41 
		& & 7.70 & 12.42 & 8.11 & 9.64 
		& &   3.16 &   3.46 &   3.14 &   3.46 
		\\ 
		& $i=   2$ 
		& & 0.83 & 2.21 & 0.83 & 1.61 
		& & 3.17 & 6.40 & 3.28 & 5.47 
		& &   3.92 &   4.35 &   3.89 &   4.34 
		\\ 
		& $i=   5$ 
		& & 0.76 & 2.60 & 0.76 & 3.54 
		& & 2.82 & 6.09 & 2.82 & 9.91 
		& &   6.38 &   7.42 &   6.41 &   7.47 
		\\ 
		& $i=  10$ 
		& & 0.80 & 2.20 & 0.80 & 1.89 
		& & 3.20 & 4.17 & 3.20 & 3.86 
		& &   8.98 &  12.14 &   9.05 &  12.18 
		\\ 
		& $i=  20$ 
		& & 0.80 & 1.36 & 0.80 & 1.35 
		& & 3.23 & 2.74 & 3.23 & 2.73 
		& &   9.26 &  20.86 &   9.32 &  20.89 
		\\ 
		& $i=  50$ 
		& & 0.80 & 1.16 & 0.80 & 1.16 
		& & 3.25 & 2.64 & 3.25 & 2.64 
		& &   9.32 &  52.85 &   9.38 &  52.89 
		\\\bottomrule
	\end{tabular*}	
	\label{tab:results_1}
\end{table*}
{The corresponding estimation errors over time can be seen in Figure~\ref{fig:example_e}, which also provides the estimation error for each suboptimal estimator using $T=t$, i.e., without re-initializing the auxiliary observer.
This obviously leads to a much worse result since the suboptimal estimators are not able to recover from the poor transient behavior of the Luenberger observer without re-initialization (cf. Remark~\ref{rem:choice_T})}, showing the effectiveness of the proposed re-initialization strategy.

{For a more detailed numerical comparison, we employ two different performance metrics: the sum-of-squared errors (SSE) and the sum-of-normed errors (SNE), defined as $\mathrm{SSE} := \sum_{i=0}^{t_{\text{sim}}}|x_i-\hat{x}_i|^2$
and $\mathrm{SNE} := \sum_{i=0}^{t_{\text{sim}}}|x_i-\hat{x}_i|$, where $t_{\text{sim}} = 60$ is the simulation length. 
To evaluate the computational complexity, we also define the average computation time per sample $\tau_{\mathrm{a}}$, considering $t\in\mathbb{I}_{\geq N}$.}
Table~\ref{tab:results_1} compares the values of SSE, SNE, and $\tau_{\mathrm{a}}$ for different configurations of the proposed suboptimal estimator and for different $i$ representing the maximum number of iterations allowed solving the respective NLP.
For $N=3$ and $T=5$, we generally find that the SSE is smaller when using the quadratic cost function, and conversely, that the SNE is smaller when using the time-discounted cost function (at least for $i \geq 15$). 
This behavior was to be expected and is clearly due to the different objectives, where we first minimize the squared decision variables and second minimize their norms.
We also directly observe that the quadratic cost function results in well-posed NLPs, where the respective optimizer is already nearly converged after $i=10$ iterations.
In contrast, the time-discounted cost function leads to more complex numerical problems{, yielding slightly larger computation times and a larger number of maximum iterations} to obtain satisfactory suboptimal results, compare Remark~\ref{rem:implementing}.
To examine the influence of longer estimation horizons, we additionally choose $N{\,=\,}10$ and $T{\,=\,}15$.
As expected, the estimation results improve as $N$ increases, while the computation time also increases.
{Overall, this shows that performing only a few iterations of the optimizer already leads to significantly better estimation results compared to the auxiliary observer, demonstrating the effectiveness of the proposed suboptimal MHE framework.}

\subsubsection*{{Comparison with fast MHE schemes \cite{Kuehl2011,Wynn2014}}}
{We compare the proposed suboptimal methods to the fast MHE schemes from \cite{Kuehl2011,Wynn2014}, which are referred to as \textit{f-MHE~1} and \textit{f-MHE~2} in remainder of this paper, respectively.
Both schemes rely on the generalized Gauss-Newton (GNN) algorithm and employ a quadratic cost function in filtering form, i.e., where the estimation horizon at a given time $t$ includes the most recent measurement $y_t$.
For a meaningful comparison, we implement the proposed suboptimal MHE scheme in a similar fashion, i.e., with the quadratic cost function \eqref{eq:costfunction} in filtering form\footnote{
	{Note that the results derived in Sections~\ref{sec:candidate_nominal}-\ref{sec:constraints} for the prediction form of MHE (i.e., without incorporating the current measurement $y_t$ at time $t$) can be straightforwardly extended to filtering MHE, compare also~\cite{Allan2019a}.}}
and using the GNN algorithm (with Lagrangian relaxation, compare~\cite{Wynn2014}) for solving the optimization problem.
We furthermore use the candidate solution based on the nominal trajectory \eqref{eq:candidate_NL_proj} and on the observer trajectory \eqref{eq:candidate_proj} and denote the resulting estimators as \textit{s-MHE~1} and \textit{s-MHE~2}, respectively.
For each estimator, we choose $N=3$ (and $T=5$ for \mbox{s-MHE~1/2}), parameterize each cost function according to the choices from above, and incorporate the constraints $\mathbb{X}$ using a simple active-set method, compare \cite{Jategaonkar2000}.}

{Table \ref{tab:comparison} shows the estimation results in terms of SSE and computation time $\tau_{\mathrm{a}}$ as defined above, averaged over 100 simulations.}
\begin{table}
	\centering
	\caption{{SSE and $\tau_{\mathrm{a}}$ for suboptimal MHE with $N{=}3\ (T{=}5)$ using cost function \eqref{eq:costfunction} and candidate solutions \eqref{eq:candidate_NL_proj} (s-MHE 1) and \eqref{eq:candidate_proj} (s-MHE 2) compared to fast MHE schemes \cite{Kuehl2011} (f-MHE 1) and~\cite{Wynn2014} (f-MHE 2) for different numbers of GNN iterations $i$.}}
	\setlength\tabcolsep{4.3pt} 
	\begin{tabular*}{\columnwidth}{@{\extracolsep{\fill}}c cc cc cc cc}
		\toprule
		\multirow{2}{*}{\ \ $i$} 
		& \multicolumn{2}{c}{s-MHE 1} 
		& \multicolumn{2}{c}{s-MHE 2}  
		& \multicolumn{2}{c}{f-MHE 1 \cite{Kuehl2011}}  
		& \multicolumn{2}{c}{f-MHE 2 \cite{Wynn2014}} 
		\\
		\cmidrule(lr){2-3}\cmidrule(lr){4-5}
		\cmidrule(lr){6-7}\cmidrule(lr){8-9}
		& SSE &  $\tau_a \ [\mathrm{ms}]$
		& SSE &  $\tau_a \ [\mathrm{ms}]$
		& SSE &  $\tau_a \ [\mathrm{ms}]$
		& SSE &  $\tau_a \ [\mathrm{ms}]$
		\\\midrule 
		\ \ 0 
		& 3.50 & 0.12
		& 2.60 & 0.12
		& 31.06 & 0.31
		& 31.06 & 0.02
		\\
		\ \ 1 
		& 0.73 & 0.59
		& 0.69 & 0.57 
		& 0.62 & 0.75
		& 0.61 & 0.44
		\\
		\ \ 2 
		& 0.50 & 0.98
		& 0.48 & 0.96 
		& 0.50 & 1.19 
		& 0.51 & 0.83 
		\\
		\ \ 5
		& 0.48 & 2.21
		& 0.47 & 2.20 
		& 0.47 & 2.50 
		& 0.47 & 2.00 
		\\\bottomrule
	\end{tabular*}	
	\label{tab:comparison}
\end{table}
{Here, it can be seen that the estimation results of s-MHE~1/2 are more accurate (in terms of SSE) and also require less computation time compared to the corresponding values in Table~\ref{tab:results_1}, which results from the fact that we use the filtering (instead of prediction) form of MHE and the GNN algorithm (instead of IPOPT), respectively.
Performing $i>0$ GNN iterations reveals that the considered estimation methods perform very similarly, both when comparing their accuracy (SSE) and their computation times ($\tau_{\mathrm{a}}$); thus, conceptually related differences on performance\footnote{
	{As expected, s-MHE 1/2 perform slightly faster than f-MHE~1 and slower than f-MHE~2, which is mainly due to the respective design of the cost function.
	More specifically, s-MHE 1/2 has a prior weighting $\Gamma$ involving constant parameters only, f-MHE 1 has a time-varying prior weighting where the parameters of $\Gamma$ needs to be updated each time-step using a QR decomposition (cf.~\cite{Kuehl2011}), and f-MHE 2 has no prior weighting and furthermore penalizes solely the fitting error (cf.~\cite{Wynn2014}).}
} are therefore rather marginal (in this setup).}
{However, our proposed suboptimal MHE framework is more flexible than that of \cite{Kuehl2011,Wynn2014}, as it is applicable to a larger class of nonlinear detectable systems and allows for a completely free choice of optimization algorithm.
Moreover, we can provide robust stability guarantees independent of the horizon length and the number of iterations performed; in contrast, fMHE 1/2 provide either no guarantees at all (cf. \cite{Kuehl2011}) or only for observable systems (cf.~\cite{Wynn2014}), heavily relying on the convergence properties of GNN.}

\section{Conclusions}
{In this paper, we presented a suboptimal MHE framework for a general class of nonlinear systems and established robust stability subject to unknown disturbances independent of: (i)~the horizon length; (ii)~the chosen optimization algorithm; (iii)~the number of solver iterations performed at each time step.}
This is crucial in order to achieve real-time applicability of MHE in cases where the optimization problem cannot be solved to optimality within one fixed sampling interval.
We considered both a standard {least squares} and a time-discounted cost function, where the former yields good performance in practice and the latter improved theoretical guarantees, {in particular}, disturbance gains that are uniformly valid for all~$N$.
The simulation example revealed that the proposed re-initialization strategy can be very effective, in particular in case of a poor transient behavior of the auxiliary observer, while constraint satisfaction of the suboptimal estimator could be guaranteed at all times.
{Moreover, with only a few iterations of the optimizer, we were able to significantly improve the estimation results of the auxiliary observer and overall achieve performance comparable to the fast MHE schemes from \cite{Kuehl2011,Wynn2014}, both in terms of accuracy and required computation time.
Future work might be based on the new Lyapunov approaches for nonlinear MHE presented in~\cite{Allan2020a,Schiller2022}.}


\begin{IEEEbiography}[{\includegraphics[width=1in,height=1.25in,clip,keepaspectratio]{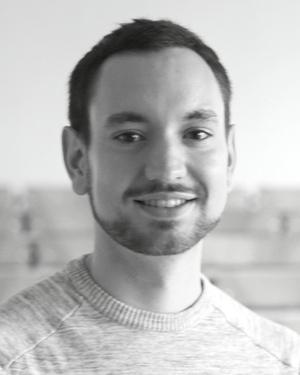}}]{Julian D. Schiller} 
	received his Master degree in Mechatronics from the Leibniz University Hannover, Germany, in 2019. 
	Since then, he has been a research assistant at the Institute of Automatic Control, Leibniz University Hannover, where he is working on his Ph.D. under the supervision of Prof. Matthias A. Müller. 
	His research interests are in the area of optimization-based state estimation and the control of nonlinear systems.
\end{IEEEbiography}

\begin{IEEEbiography}[{\includegraphics[width=1in,height=1.25in,clip,keepaspectratio]{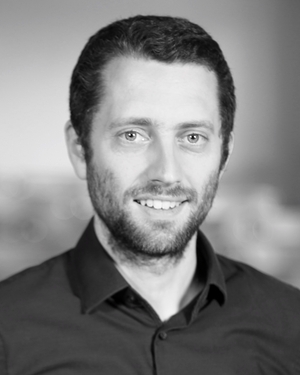}}]{Matthias A. Müller} 
	received a Diploma degree in Engineering Cybernetics from the University of Stuttgart, Germany, and an M.S. in Electrical and Computer Engineering from the University of Illinois at Urbana-Champaign, US, both in 2009. 
	In 2014, he obtained a Ph.D. in Mechanical Engineering, also from the University of Stuttgart, Germany, for which he received the 2015 European Ph.D. award on control for complex and heterogeneous systems. 
	Since 2019, he is director of the Institute of Automatic Control and full professor at the Leibniz University Hannover, Germany. He obtained an ERC Starting Grant in 2020 and is recipient of the inaugural Brockett-Willems Outstanding Paper Award for the best paper published in Systems \& Control Letters in the period 2014-2018. 	
	His research interests include nonlinear control and estimation, model predictive control, and data-/learning-based control, with application in different fields including biomedical engineering.
\end{IEEEbiography}

\end{document}